%% file: SUS-23-012_temp.tex
\pdfoutput=1
\documentclass[11pt,twoside,a4paper,cmspaper,final,collab]{cms-tdr}
\def\svnVersion{0672ad3}\def\svnDate{2025/10/31}\def\cmsCernNoTag{CERN-EP-2025-100}\def\cmsCernDate{\today}\def\cmsMessage{Published in the Journal of High Energy Physics as \href{http://dx.doi.org/10.1007/JHEP10(2025)170}{\doi{10.1007/JHEP10(2025)170}.}}
\begin{document}\cmsNoteHeader{SUS-23-012}

\newcommand{\Ppp}{\ensuremath{\Pp\Pp}\xspace}
\newcommand{\dmh}{{DM+h}\xspace}
\newcommand{\baryonicZprime}{baryonic {\ensuremath{\PZpr}}\xspace}
\newcommand{\sqrtthirteentev}{\ensuremath{\sqrt{s}=13\TeV}\xspace}
\newcommand{\Htt}{\ensuremath{\Ph\to\\PGtp\PGtm}\xspace}
\newcommand{\wjet}{{\PW{}+jets}\xspace}
\newcommand{\zjet}{{\PZ{}+jets}\xspace}
\newcommand{\mttot}{\ensuremath{M_\mathrm{T}^{\text{tot}}}\xspace}
\newcommand{\mchi}{\ensuremath{m_{\PGc}}\xspace}
\newcommand{\ychi}{\ensuremath{y_{\PGc}}\xspace}
\newcommand{\htt}{\ensuremath{\Ph\to\PGt\PGt}\xspace}
\newcommand{\mtt}{\ensuremath{m_{\PGt\PGt}}\xspace}
\newcommand{\mZpr}{\ensuremath{m_{\PZpr}}\xspace}
\newcommand{\ma}{\ensuremath{m_{\Pa}}\xspace}
\newcommand{\mA}{\ensuremath{m_{\PSA}}\xspace}
\newcommand{\mh}{\ensuremath{m_{\Ph}}\xspace}
\newcommand{\mH}{\ensuremath{m_{\PH}}\xspace}
\newcommand{\mHpm}{\ensuremath{m_{\PHpm}}\xspace}
\newcommand{\sintheta}{\ensuremath{\sin\theta}\xspace}
\newcommand{\costheta}{\ensuremath{\cos\theta}\xspace}
\newcommand{\lambdathree}{\ensuremath{\lambda_{3}}\xspace}
\newcommand{\lambdaPone}{\ensuremath{\lambda_{P_{1}}}\xspace}
\newcommand{\lambdaPtwo}{\ensuremath{\lambda_{P_{2}}}\xspace}
\newcommand{\twohdma}{{2HDM+a}\xspace}
\newcommand{\cosbetaminusalpha}{\ensuremath{\cos(\beta-\alpha)}\xspace}
\newcommand{\sinbetaminusalpha}{\ensuremath{\sin(\beta-\alpha)}\xspace}
\newcommand{\SigCL}{\ensuremath{\sigma_{\text{95\% \,\CL}}}\xspace}
\newcommand{\PhB}{\ensuremath{\Ph_{\mathrm{B}}}\xspace}
\newcommand{\Uone}{\ensuremath{U(1)_{\mathrm{B}}}\xspace}
\newcommand{\pte}{\ensuremath{\pt^{\Pe}}\xspace}
\newcommand{\ptvece}{\ensuremath{\vec{p}_{\mathrm{T}}^{\Pe}}\xspace}
\newcommand{\pttauh}{\ensuremath{\pt^{\tauh}}\xspace}
\newcommand{\ptmu}{\ensuremath{\pt^{\PGm}}\xspace}
\newcommand{\ptvecmu}{\ensuremath{\vec{p}_{\mathrm{T}}^{\PGm}}\xspace}
\newcommand{\ptl}{\ensuremath{\pt^{\Pell}}\xspace}
\newcommand{\ptvecl}{\ensuremath{\vec{p}_{\mathrm{T}}^{\Pell}}\xspace}
\newcommand{\emu}{\ensuremath{\Pe\PGm}\space}
\newcommand{\etau}{\ensuremath{\Pe\tauh}\xspace}
\newcommand{\mutau}{\ensuremath{\PGm\tauh}\xspace}
\newcommand{\tautau}{\ensuremath{{\PGt\PGt}}\xspace}
\newcommand{\tauhtauh}{\ensuremath{\tauh\tauh}\xspace}
\cmsNoteHeader{SUS-23-012}
\title{Search for dark matter produced in association with a Higgs boson decaying to a \texorpdfstring{\PGt}{tau} lepton pair in proton-proton collisions at \texorpdfstring{$\sqrt{s}=13\TeV$}{sqrt(s)= 13 TeV}}

\date{\today}

\abstract{
A search for dark matter particles produced in association with a Higgs boson decaying into a pair of \PGt leptons is performed using data collected in proton-proton collisions at a center-of-mass energy of 13\TeV with the CMS detector. The analysis is based on a data set corresponding to an integrated luminosity of 101\fbinv collected in 2017--2018. No significant excess over the expected standard model background is observed. This result is interpreted within the frameworks of the 2HDM+a and baryonic \PZpr benchmark simplified models. The 2HDM+a model is a type-II two-Higgs-doublet model featuring a heavy pseudoscalar with an additional light pseudoscalar. Upper limits at 95\% confidence level are set on the product of the production cross section and the branching fraction for each of these two simplified models. Heavy pseudoscalar boson masses between 400 and 700\GeV are excluded for a light pseudoscalar mass of 100\GeV. For the baryonic \PZpr model, a statistical combination is made with an earlier search based on a data set of 36\fbinv collected in 2016. In this model, \PZpr boson masses up to 1050\GeV are excluded for a dark matter particle mass of 1\GeV.
}

\hypersetup{%
pdfauthor={CMS Collaboration},%
pdftitle={Search for dark matter produced in association with a Higgs boson decaying to a tau lepton pair in proton-proton collisions at sqrt(s) = 13 TeV},%
pdfsubject={CMS},%
pdfkeywords={CMS, dark matter, Higgs boson, tau lepton, mono-X}}

\maketitle

\section{Introduction}\label{sec:intro}

Evidence for dark matter (DM) and its influence on the observed large-scale structure of the universe is well supported by numerous studies~\cite{Feng:2010gw,Porter:2011nv,Clowe:2003tk,Planck:2015fie}. This evidence, along with the hypothesis that DM consists of long-lived weakly interacting massive particles, drives the search for potential DM production at the CERN LHC. Given the stringent constraints on weak interactions between DM and standard model (SM) particles, models incorporating beyond-the-SM (BSM) scalar sectors are of significant interest.

If DM is produced at the LHC, it is likely to escape direct detection because of its rare interaction with the matter that makes up the detectors. However, if DM candidates are produced alongside SM particles, such as the 125\GeV Higgs boson, they may create an imbalance in recorded momentum, resulting in a large amount of missing transverse momentum (\ptmiss). Such signatures are generically called ``mono-Higgs". This paper reports on a search for the production of DM particles (\PGc) recoiling against a Higgs boson, which is reconstructed through its decay into a pair of \PGt leptons in both leptonic and hadronic decays. The data set used in this analysis was collected from proton-proton (\Ppp) collisions at a center-of-mass energy of 13\TeV by the CMS experiment in 2017 and 2018, corresponding to an integrated luminosity of 101\fbinv. A statistical combination is performed considering a model probed by an earlier search, conducted at the same center-of-mass energy with 36\fbinv of data collected in 2016~\cite{CMS:2018nlv}.

\begin{figure}[hbpt]
  \centering
  \includegraphics[width=0.4\textwidth]{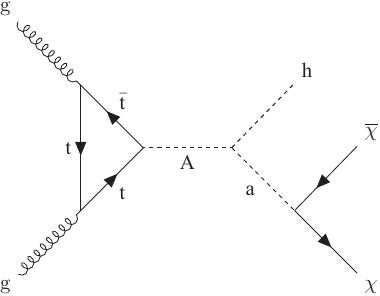}\hspace{2em}
  \includegraphics[width=0.35\textwidth]{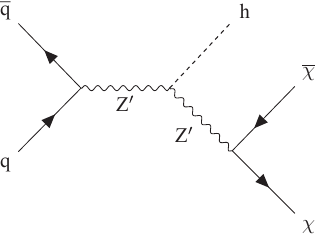}
	\caption{Representative Feynman diagrams for leading order DM-associated production of a Higgs boson in the \twohdma (left) and \baryonicZprime (right) models.}
  \label{fig:feyndiag}
\end{figure}

The ``mono-Higgs" signature arises from new interactions between DM and SM particles. Several models suggest enhanced couplings between BSM particles and scalar bosons~\cite{Bertone:2004pz,Burgess:2000yq,March-Russell:2008lng}. Although the Higgs boson could theoretically be radiated from initial-state quarks, this process is highly suppressed because of small Yukawa couplings or loop processes~\cite{Carpenter:2013xra, Berlin:2014cfa, Petrov:2013nia}. This analysis considers two simplified benchmark models for \dmh production~\cite{Abercrombie:2015wmb}. The \twohdma model~\cite{Abe:2018bpo, Bauer:2017ota} is a canonical two Higgs doublet model with an additional pseudoscalar singlet interacting with DM particles. This model features five Higgs bosons: one SM-like Higgs boson (\Ph), two charged Higgs bosons (\PHpm), one heavier neutral scalar (\PH), and one heavier neutral pseudoscalar (\PSA), as well as an additional pseudoscalar (\Pa). The Feynman diagram describing DM production in this model is shown in Fig.~\ref{fig:feyndiag} (left). The \baryonicZprime model~\cite{Carpenter:2013xra} predicts a new massive vector mediator \PZpr that can emit a Higgs boson and then decay to a pair of Dirac fermionic DM particles, leading to the Feynman diagram shown in Fig.~\ref{fig:feyndiag} (right).

According to Ref.~\cite{Abercrombie:2015wmb}, a benchmark scenario is defined by fixing several model parameters based on both electroweak and flavor measurements and phenomenological considerations, such as the requirement that the Higgs potential is stable. Assuming the alignment of the \twohdma model and the decoupling limits, \mh is set to 125\GeV. From electroweak precision constraints, the masses of the heavier Higgs bosons (\PHpm, \PH, \PSA) are set to be equal, leaving five free parameters of the model: the mass of the light pseudoscalar \ma, the mass of the heavy pseudoscalar \mA, the mass of the DM candidate, \mchi, \sintheta, where $\theta$ is the mixing angle of the two CP-odd weak spin-0 eigenstates, and the ratio of the vacuum expectation values of the two Higgs doublets, \tanb. The variations of the \sintheta and \tanb parameters change the signal strength and have a prominent impact on the shapes of the \ptmiss distribution of the mono-Higgs production.

In the baryonic \PZpr model~\cite{Carpenter:2013xra}, the baryonic gauge boson \PZpr arises from a new \Uone baryon number symmetry. A baryonic Higgs boson (\PhB) is introduced to break this symmetry spontaneously, generating the \PZpr boson mass via a coupling dependent on the \PhB vacuum expectation value. The \PZpr boson couplings to quarks and DM particles are proportional to the \Uone gauge couplings. There is a mixing between \PhB and the 125\GeV Higgs boson (\Ph), allowing the \PZpr to radiate an h. The coupling strength between the \PZpr boson and \Ph is $g_{h\PZpr\PZpr}$ = $(m_{\PZpr}^{2} \sin \zeta)/v_B$, where the parameter $\sin \zeta$ governs the mixing angle between the \PhB and \Ph, and $v_B$ is the vacuum expectation value of \PhB. In this model, stable baryonic states are the candidate DM particles. This paper considers a \PZpr resonance with a mass between 10 and 2500\GeV, and DM particle masses between 1 and 800\GeV. As suggested in Ref.~\cite{Boveia:2016mrp}, the mediator-DM coupling is fixed to 1, and the mediator-quark coupling to 0.25 to avoid the constraints from dijet searches at the LHC. 

Although the branching ratio of the SM Higgs boson to two \PGt leptons is lower than that to \bbbar, this analysis takes advantage of several distinct benefits of the h that decays to a \PGt pair. It benefits from small SM backgrounds and does not rely on \ptmiss triggers. As a result, the $\PGt\PGt$ channel serves as a complementary approach to the bb channel, enabling the exploration of dark matter scenarios with lower \ptmiss values.

SM processes can also give rise to a signature with \ptmiss and two \PGt leptons consistent with the decay products of a Higgs boson. One such important SM process is the production of a Higgs boson in association with a \PZ boson that decays into neutrinos. Although this process has a relatively low cross section, it may produce final states indistinguishable from the signal. Other SM processes that can result in final states with \ptmiss and two \PGt lepton decays include the production of a \PZ boson, a pair of vector bosons ($\PW\PW$, $\PW\PZ$, $\PZ\PZ$), a top quark-antiquark pair, or a Higgs boson in association with a \PW boson. Separating the signal from these processes is challenging because of their higher cross sections, finite detector resolution, acceptance limitations, and the inherent uncertainty introduced by the presence of two neutrinos, which can potentially mimic signal events.

The previous search in the same final state by the CMS experiment~\cite{CMS:2018nlv} was based on data collected at a center-of-mass energy of 13\TeV and corresponding to an integrated luminosity of 36\fbinv. These results were interpreted in terms of the baryonic \PZpr model. A similar search was conducted by the ATLAS experiment at a center-of-mass energy of 13\TeV, using a data set corresponding to an integrated luminosity of 139\fbinv~\cite{ATLAS:2023ild}. The results of this search were interpreted in terms of a \twohdma model featuring two scalar Higgs doublets and a pseudoscalar singlet field. No significant excess over the SM prediction was observed in either study.

The analysis reported here benefits from a larger data set and employs a method based on control samples in data, known as the ``misidentification factor (MF) method'', described in Refs.~\cite{CMS:2018rmh,CMS:2018lkr}. This method collectively estimates background events with quark- or gluon-jets misidentified as hadronically decaying \PGt leptons ($\tauh$), instead of measuring this background separately for $\PW$+jets and multijet events arising from quantum chromodynamics (QCD).

The remainder of this paper is organized as follows. In Section~\ref{sec:cms}, the CMS detector and the reconstruction of the objects and event variables used in the analysis are introduced. Section~\ref{sec:dataset} details the data set and the Monte Carlo (MC) simulated samples used in the analysis. Section~\ref{sec:strategy} describes the event selection and analysis strategy for the study, while Section~\ref{sec:sys_unc} outlines the systematic uncertainties affecting the analysis. Finally, Section~\ref{sec:results} presents the results and their interpretation, and a summary is provided in Section~\ref{sec:summary}. The tabulated results are also provided in the HEPData record~\cite{hepdata}.

\section{The CMS detector and event reconstruction}\label{sec:cms}

The central feature of the CMS apparatus is a superconducting solenoid of 6\unit{m} internal diameter, providing a magnetic field of 3.8\unit{T}. Within the solenoid volume are a silicon pixel and strip tracker, a lead tungstate crystal electromagnetic calorimeter (ECAL), and a brass and scintillator hadron calorimeter (HCAL), each composed of a barrel and two endcap sections. Forward calorimeters extend the pseudorapidity coverage provided by the barrel and endcap detectors. Muons are detected in gas-ionization chambers embedded in the steel flux-return yoke outside the solenoid.

Events of interest are selected using a two-tiered trigger system. The first level (L1), composed of custom hardware processors, uses information from the calorimeters and muon detectors to select events at a rate of around 100\unit{kHz} within a fixed latency of about 4\mus~\cite{Sirunyan:2020zal}. The second level, known as the high-level trigger, consists of a farm of processors running a version of the full event reconstruction software optimized for fast processing, and reduces the event rate to a few \unit{kHz} before data storage~\cite{CMS:2016ngn,CMS:2024aqx}. A more detailed description of the CMS detector, together with a definition of the coordinate system used and the relevant kinematic variables, can be found in Refs.~\cite{Chatrchyan:2008zzk,CMS:2023gfb}.

The particle-flow (PF) algorithm~\cite{CMS-PRF-14-001} combines information from all the different parts of the CMS detector to identify and reconstruct individual particles. These particles, called PF candidates, can be charged or neutral hadrons, electrons, photons, or muons. The reconstructed track vertex with the largest value of summed physics-object $\pt^2$, where \pt is the transverse momentum of the objects, is taken to be the primary vertex (PV) of the \Ppp interaction. Any other collision vertices in the event are associated with additional mostly soft inelastic \Ppp collisions called pileup (PU). In the 2017--2018 data sets the average number of interactions per bunch crossing was 32.

Electron candidates are reconstructed by combining clusters of energy deposits in the ECAL with hits in the silicon tracker~\cite{CMS:2020uim}. The momentum resolution for electrons with $\pt \approx 45\GeV$ from $\PZ \to \Pe \Pe$ decays ranges from 1.6 to 5\%. Electron identification is performed using a multivariate discriminant derived from track quality, shower shape, and kinematic quantities. For this analysis, a medium working point with an identification efficiency of 90\% is used, with a misidentification rate of $\approx$1\% from jets in the relevant kinematic region~\cite{CMS:2020uim}. Muons are measured using detection planes that employ three technologies: drift tubes, cathode strip chambers, and resistive-plate chambers~\cite{CMS:2018rym}. Muons in the event are reconstructed by performing a simultaneous track fit to hits in the tracker and in the muon chambers. The momentum resolution, for muons with $\pt$ up to $100\GeV$, is 1\% in the barrel and 3\% in the endcaps. The $\pt$ resolution in the barrel is better than 7\% for muons with $\pt$ up to $1\TeV$. For this study, muon identification requirements with an efficiency of approximately 99\% are chosen, with a misidentification rate below 0.2\% for pions~\cite{CMS:2020uim}.

The contribution of backgrounds with misidentified electrons or muons can further be minimized by isolating the corresponding lepton candidate from any hadronic activity. The isolation variable, defined by the following Eq. (\ref{eq:isolation}), is used:
\begin{equation}
I^{\Pell}_{\text{rel}} = \frac{1}{\pt^{\Pell}} \left[ \Sigma\pt^{\text{charged}} + \text{max}\left( 0, \Sigma\ET^{\text{neutral}} + \Sigma\ET^{\PGg} - \pt^{\text{PU}}\right) \right],
\label{eq:isolation}
\end{equation}
where lepton $\Pell$ refers to either an electron or muon, $\pt^{\Pell}$ corresponds to the lepton \pt, $\Sigma\pt^{\text{charged}}$ corresponds to the \pt sum of all charged particles, and $\Sigma\ET^{\text{neutral}}$ and $\Sigma\ET^{\PGg}$ correspond to the transverse energy \ET sum of neutral hadrons and photons, respectively. This isolation is predefined in a cone of radius $\Delta R = \sqrt{\smash[b]{(\Delta\eta)^2 + (\Delta\phi)^2}}$ around the lepton direction at the PV, where $\Delta\eta$ and $\Delta\phi$ (measured in radians) correspond to the angular distances of the particle to the lepton in the $\eta$ and $\phi$ directions, respectively. The cone sizes for electrons and muons are $\Delta R=0.3$ and 0.4, respectively. The lepton itself is excluded from the isolation calculation. To mitigate the contamination from PU, only those charged particles whose tracks are associated with the PV are taken into account. For neutral hadrons and photons, it is hard to determine whether they originate from the PV or from PU. To account for this, an estimated contribution from PU ($\pt^{\text{PU}}$) is subtracted from the total measured energy of neutral hadrons and photons. For electrons, this is estimated from the mean energy flow per unit area, while for muons it is obtained from tracks not associated with the PV. The isolation criteria mentioned in Section~\ref{sec:selection} have an efficiency for isolated electrons and muons from \PGt lepton decays well above 95\%.

For each event, hadronic jets are reconstructed from the PF candidates using the infrared- and collinear-safe anti-\kt algorithm~\cite{Cacciari:2008gp, Cacciari:2011ma} with a distance parameter R of 0.4. Jet momentum is determined as the vector sum of all particle momenta in the jet, and is found from simulation to be, on average, within 5 to 10\% of the true momentum over the entire \pt spectrum and detector acceptance~\cite{Khachatryan:2016kdb}. In 2017, the ECAL endcaps were subject to an increased noise level affecting the reconstruction of jets. Jets with $\pt < 50\GeV$ in the range of $2.65 < \abs{\eta} < 3.14$ have been excluded from the analysis. Jets resulting from the hadronization of b quarks are used to separate signal from top quark-antiquark pair (\ttbar) events. These are tagged by exploiting the neural net based \textsc{DeepCSV} algorithm as described in Refs.~\cite{CMS:2017wtu,Bols:2020bkb}. The medium working point for \textsc{DeepCSV} is chosen, which achieves approximately 66\% efficiency with a 1\% misidentification probability. Jets with $\pt > 30\GeV$ and $\abs{\eta} < 4.7$ and \PQb jets with $\pt > 20\GeV$ and $\abs{\eta} < 2.5$ are used.

Jet energy corrections are derived from simulation to bring the measured response of jets to that of particle-level jets on average. In situ measurements of the momentum balance in dijet, $\PGg$+jet, $\PZ$+jet, and multijet events are used to account for any residual differences in the jet energy scale (JES) and jet energy resolution (JER) in data and simulation~\cite{Khachatryan:2016kdb}.  The JER amounts typically to 15--20\% at 30\GeV, 10\% at 100\GeV, and 5\% at 1\TeV~\cite{Khachatryan:2016kdb}. Additional selection criteria~\cite{CMS-PAS-JME-16-003} are applied to each jet to remove jets potentially dominated by anomalous contributions from various subdetector components or reconstruction failures.

Hadronic \PGt lepton decays (\tauh) are reconstructed from jets using the hadrons-plus-strips algorithm~\cite{CMS:2018jrd}, which combines one or three tracks with energy deposits in the calorimeters to identify the decay modes. Neutral pions are reconstructed by clustering the PF electrons and photons falling in $\eta$-$\phi$ regions, called the strips, with a dynamic size in $\eta$-$\phi$, where the strip size varies as a function of the $\pt$ of the electron or photon candidate. To distinguish genuine \tauh decays from jets originating from the hadronization of quarks or gluons, and from electrons, or muons, the \textsc{DeepTau} algorithm is used~\cite{CMS:2022prd}. Information from all individual reconstructed particles near the \tauh axis is combined with properties of the \tauh candidate and the event.
The input variables to the neural net include variables related to the $\tauh$, four-momentum, charge, lifetime, isolation, the compatibility of the $\tauh$ track with the PV, observables associated with the $\eta$ and $\phi$ distributions of the energy reconstructed in the strips, and other variables. The thresholds on the output discriminants~\cite{CMS:2022prd}, dependent on $\pt$, are chosen to define several working points, which provide good $\tauh$ identification efficiencies. Table~\ref{tab:deepTau} shows the target \tauh identification efficiences for the different working points that are used in the analysis to discriminate against electrons ($D_\Pe$), muons ($D_\PGm$), and jets ($D_\text{jet}$). The probability of a jet to be misidentified as a \tauh candidate by the \textsc{DeepTau} algorithm depends on the \pt and quark flavor of the jet. In simulated events from \PW boson production in association with jets, it has been estimated to be 0.43\% for a genuine \tauh identification efficiency of 70\%. The misidentification rate for electrons (muons) is 2.60 (0.03)\% for a genuine \tauh identification efficiency of 80 ($>$99)\%.

\begin{table}[htbp]
  \centering
  \caption{Target \tauh identification efficiencies for the different working points defined for the three different discriminants that are used in the analysis. These identification efficiencies are evaluated for genuine \tauh candidates in simulated $\PH\to\PGt\PGt$ events with $\pt \in [30, 70]\GeV$ for \tauh. 
    \label{tab:deepTau}
  }
  \begin{tabular}{lllll}
    & Tight    & Medium  & VLoose  & VVVLoose \\
    \hline
    $D_\Pe$ &  80\%    & \NA      & \NA    & 99.5\% \\
    $D_\PGm$ &  99.5\%  & \NA      & 99.95\% & \NA  \\
    $D_\text{jet}$ &  \NA    & 70\%      & \NA    & 98\% \\
  \end{tabular}
\end{table}

The negative vector sum of the transverse momenta ($\ptvecmiss$) is computed from all the PF candidates in an event, and its magnitude is denoted as \ptmiss~\cite{Sirunyan:2019kia}. The value of \ptvecmiss is adjusted to account for corrections to the energy scale and resolution of the reconstructed jets in the event.  Anomalous high-\ptmiss events can arise from various issues such as reconstruction failures, detector malfunctions, or non-collision backgrounds. These events are rejected by dedicated filters that are designed to eliminate more than 85--90\% of the spurious high-\ptmiss events with a signal efficiency exceeding 99.9\%~\cite{Sirunyan:2019kia}. The \ptmiss is used for the discrimination of \PW boson production in association with jets ($\PW$+jets) from signal by exploiting the transverse mass:
\begin{equation}
    \mT^{\Pell} = \sqrt{2\pt^{\Pell}\ptmiss(1-\cos\Delta\phi)},
\label{eq:transversemass}
\end{equation}
where $\Pell$ stands for an electron or a muon, and $\Delta\phi$ is the azimuthal angular separation between $\pt^{\Pell}$ and \ptmiss.

\section{Data and simulated samples}\label{sec:dataset}

The analysis is performed with \Ppp collision data at \sqrtthirteentev collected with the CMS detector in 2017 and 2018. The data correspond to an integrated luminosity of 101\fbinv. The analysis strategy and event selection were optimized using simulated samples of associated DM+\Ph(\tautau) production via the two benchmark models discussed in Section~\ref{sec:intro}. The \MGvATNLO~v2.6.0~\cite{Alwall:2014hca} generator is used to generate both the \twohdma and \baryonicZprime signals at leading order (LO). The decay of the SM-like Higgs boson is simulated by \PYTHIA~8.205~\cite{Sjostrand:2014zea} using the CP2 tune~\cite{CMS:2019csb}.

The simulation is performed for mass values scanning the available phase space in the (\ma,\mA) plane, for the \twohdma model (see Table~\ref{tab:signal_2hdma}), and in the (\mchi,\mZpr) plane for the \baryonicZprime model (see Table~\ref{tab:signal_zpb}).

\begin{table}[htbp]
    \topcaption{Mass values scanning the available phase space in the (\ma,\mA) plane, used in the simulation for the \twohdma model. The phase space is limited by $\mA - \ma < 125\GeV$, where 125\GeV is the Higgs mass.}
    \label{tab:signal_2hdma}
    \centering
    \begin{tabular}{rllllllllll}

\multicolumn{1}{r}{\mA [\GeVns{}]} & \multicolumn{10}{c}{ \ma [\GeVns{}]}\\
&\multicolumn{10}{c}{ \tanb = 1, \sintheta = 0.35, \mchi = 10\GeV} \\ \hline
225  & 100 &     &     &     &     &     &     &     &     &     \\
250  &     & 125 &     &     &     &     &     &     &     &     \\
300  & 100 &     & 150 &     &     &     &     &     &     &     \\
400  & 100 &     & 150 & 200 & 250 &     &     &     &     &     \\
500  &     &     & 150 & 200 & 250 & 300 &     & 375 &     &     \\
600  & 100 &     & 150 & 200 & 250 & 300 & 350 &     & 400 & 500 \\
700  & 100 &     &     & 200 & 250 & 300 & 350 &     & 400 &     \\
800  & 100 &     & 150 & 200 & 250 & 300 & 350 &     &     & 500 \\
900  & 100 &     & 150 & 200 & 250 & 300 & 350 &     & 400 & 500 \\
1000 & 100 &     & 150 &     & 250 &     & 350 &     &     &     \\
1200 &     &     & 150 &     & 250 &     & 350 &     &     &     \\
\end{tabular}
\end{table}

\begin{table}[htbp]
    \topcaption{Mass values scanning the available phase space in the (\mchi,\mZpr) plane, used in the simulation for the \baryonicZprime model. The phase space is limited by $\mZpr - 2\mchi < 125\GeV$, where 125\GeV is the Higgs mass, and by $\mchi < \mZpr/2$.}
    \label{tab:signal_zpb}
    \centering
    \begin{tabular}{rllllllllllll}

\multicolumn{1}{r}{\mZpr [\GeVns{}]}&\multicolumn{12}{c}{ \mchi [\GeVns{}] } \\ \hline
200  & 1   &     &      &      &      &      &      &      &      &      &      &      \\
300  & 1   & 50  &      &      &      &      &      &      &      &      &      &      \\
350  &     &     & 100  &      &      &      &      &      &      &      &      &      \\
500  & 1   & 50  & 100  &      &      &      &      &      &      &      &      &      \\
650  &     & 50  & 100  & 150  &      &      &      &      &      &      &      &      \\
700  &     &     &      &      &      & 225  &      &      &      &      &      &      \\
750  &     &     &      &      &      & 225  &      &      &      &      &      &      \\
800  &     & 50  &      & 150  & 200  &      & 250  & 300  & 325  &      &      &      \\
900  &     &     &      & 150  &      &      &      &      &      &      &      &      \\
1000 & 1   & 50  & 100  &      &      &      &      &      &      &      &      &      \\
1100 &     & 50  &      &      &      &      &      &      &      &      &      &      \\
1500 & 1   & 50  & 100  &      & 200  &      &      &      &      & 400  & 600  &      \\
2000 & 1   &     & 100  &      & 200  &      &      &      &      & 400  &      & 800  \\
2500 & 1   &     &      &      &      &      &      &      &      &      &      &      \\
\end{tabular}
\end{table}

The largest backgrounds are $\PZ/\PGg^* \to \Pell\Pell/\PGt\PGt$, \wjet, $\ttbar$, single top quark, and multiboson processes. The \MGvATNLO~v2.4.2 generator is used for the $\PZ/\PGg^* \to \Pell\Pell/\PGt\PGt$ and \wjet processes, 
which are generated at LO in perturbative QCD with the MLM jet matching and merging scheme~\cite{Alwall:2007fs}. The $\PZ/\PGg^* \to \Pell\Pell/\PGt\PGt$ background is normalized to the cross section calculated at next-to-next-to-LO (NNLO) precision~\cite{Melnikov:2006kv} and is corrected for differences in the dilepton invarient mass ($m^{\Pell\Pell/\PGt\PGt}$) and transverse momentum ($\pt^{\Pell\Pell/\PGt\PGt}$) distributions using dimuon events in data~\cite{Sirunyan:2017khh}. 

The reducible SM backgrounds with genuine Higgs bosons are produced via different processes: gluon-gluon fusion ($\Pg\Pg\Ph$), vector boson fusion (VBF), in association with a vector boson ($\PV\Ph$), or in association with a pair of top quarks ($\ttbar$\Ph). These Higgs boson samples are all generated using \POWHEG 2.0~\cite{Frixione:2007vw, Alioli:2010xd, Alioli:2010xa, Alioli:2008tz}  at next-to-LO (NLO) in perturbative QCD. The $\ttbar$ and single top quark processes are generated at NLO with the \POWHEG 2.0 generator. The FxFx~\cite{Frederix:2012ps} merging scheme is used to generate diboson backgrounds with the \MGvATNLO generator at NLO. 

All simulated samples mentioned above use the NNPDF 3.1 NNLO parton distribution function (PDF) sets~\cite{NNPDF:2017mvq,Skands:2014pea} with the perturbative order matching that used in the matrix element calculations. For parton showering and hadronization, as well as for $\PGt$ lepton decays, the samples are interfaced with \PYTHIA using the CP5 tune~\cite{CMS:2019csb}.  The MC samples are processed through a full simulation of the CMS detector based on \GEANTfour~\cite{Agostinelli:2002hh} and are reconstructed with the same algorithms that are used for data. Minimum-bias collision events generated with \PYTHIA are added to the simulated samples to reproduce the PU effects in the data.

\section{Analysis strategy}\label{sec:strategy}

The search  is conducted by categorizing events on the basis of the decay modes of the \PGt leptons. This analysis focuses on the three \PGt lepton pair final states with the highest branching fractions: \etau, \mutau, and \tauhtauh. In the \etau and \mutau channels, one \PGt lepton decays leptonically to an electron or a muon along with two neutrinos, while the other \PGt lepton decays hadronically (\tauh) with one neutrino. In the \tauhtauh channel, both \PGt leptons decay hadronically, each with one neutrino. The \emu, $\Pe\Pe$, and $\PGm\PGm$ final states are excluded because of the low branching fraction of \tautau pairs to purely leptonic final states. Additionally, the $\Pe\Pe$ and $\PGm\PGm$ final states are heavily dominated by the Drell--Yan production of lepton pairs. The search is performed separately for the 2017 and 2018 data sets and then combined statistically for the two data-taking years, as well as with the published 2016 results~\cite{CMS:2018nlv}.

\subsection{Event selection}\label{sec:selection}

To select events in the \etau and \mutau channels, triggers based on the presence of a single electron or muon are employed. Additionally, in the \etau channel, a trigger relying on the presence of an electron and a \tauh candidate is also used. For the \tauhtauh channel, the triggers require the presence of two isolated \tauh objects. Each offline reconstructed \tauh candidate must be matched to a \tauh candidate at the trigger level with a $\Delta R$ separation of less than 0.5.
The efficiency of the online selection generally exceeds 90\% after the offline selection described below, and it does not exhibit strong kinematic dependencies for the leptons selected for the offline analysis, as verified with independent triggers. Events in each channel must satisfy the respective offline selection criteria for \pt, $\eta$, and $I_\text{rel}^{\Pe(\PGm)}$, as described in Table~\ref{tab:offline_selection}. These criteria are chosen to be efficient with respect to the online selection.

For the \etau channel, events are selected based on the presence of at least one electron and one \tauh candidate. The electron must satisfy a multivariate identification criterion with 90\% efficiency~\cite{CMS:2020uim}. It is also required to be isolated from any hadronic activity in the detector, with a relative isolation requirement of $I_{\text{rel}}^{\Pe} < 0.15$.  Additionally, the electron and \tauh candidate must have opposite charge and be separated by $\Delta R > 0.5$ in the $\eta$-$\phi$ plane. The \tauh candidate must pass the ``Medium'' \textsc{DeepTau} discriminant against jets. To reduce the contribution from background processes where an electron or muon is misidentified as a \tauh, the selected \tauh candidate must satisfy the anti-lepton discriminators, ``Tight'' against electrons, and ``VLoose'' against muons, as detailed in Table~\ref{tab:deepTau}.

\begin{table}[htbp]
\centering
\caption{Offline selection requirements applied to \Pe, \PGm, and \tauh candidates used for the selection of \PGt pairs. The expressions ``first lepton" and ``second lepton" refer to the lepton order of the final state in the first column. 
\label{tab:offline_selection}
}
\begin{tabular}{llll}
Final state &  Observable          & First lepton  & Second lepton\\ 
\hline
\etau       & \pt        $>$        &   25\GeV        & 30\GeV \\
            & \abs{\eta} $<$        &   2.1       & 2.3 \\
            & $I^{\Pe}_\text{rel}<$  &  0.15        & \NA \\

\mutau      & \pt        $>$        &    29\GeV        & 30\GeV \\
            & \abs{\eta} $<$        &    2.4       & 2.3 \\
            & $I^{\PGm}_\text{rel}<$  &   0.15       &\NA \\

\tauhtauh   & \pt        $>$        &    55\GeV        & 45\GeV \\
            & \abs{\eta} $<$        &    2.1       & 2.1 \\
\end{tabular}
\end{table}

Similarly, for the \mutau channel, events are selected based on the presence of at least one muon and one \tauh candidate. The muon must satisfy a PF-based identification criterion with $\approx$95\% efficiency~\cite{CMS:2018rym}. It is also required to be isolated from any hadronic activity in the detector, with a relative isolation requirement of $I_{\text{rel}}^{\PGm} < 0.15$.  Additionally, the muon and \tauh candidate must have opposite charge and be separated by $\Delta R > 0.5$ in the $\eta$-$\phi$ plane. The \tauh candidate must pass the ``Medium'' \textsc{DeepTau} discriminant against jets, the ``VVVLoose'' discriminant against electrons, and the ``Tight'' discriminant against muons, as detailed in Table~\ref{tab:deepTau}. 

Two types of vetoes are implemented in the \etau and \mutau channels to reject backgrounds. First, events containing \PQb-tagged jets with $\pt > 25\GeV$ and $\abs{\eta}<2.4$ are vetoed to suppress contributions from \ttbar and single top quark processes. Second, to mitigate contributions from backgrounds with multiple leptons, events containing additional electrons or muons, each with $\pt>10\GeV$ and $\abs{\eta}<2.4$, aside from the \PGt candidates, are excluded from further analysis. The additional electron or muon must satisfy similar identification criteria as the leading lepton but without the isolation requirement. 

In the \tauhtauh channel, events are selected based on the presence of two oppositely charged \tauh leptons. The two \tauh candidates must be separated in the $\eta$-$\phi$ plane by $\Delta R > 0.5$. Both the \tauh candidates must pass the ``Medium'' \textsc{DeepTau} discriminant against jets, the ``VVVLoose'' discriminant against electrons, and the ``VLoose'' discriminant against muons, as detailed in Table~\ref{tab:deepTau}.

In addition, for all three \PGt pair categories, \ptmiss is required to be greater than 105\GeV and the visible \pt of the $\PGt\PGt$ system must exceed 65\GeV. These stringent criteria reduce the need for isolation in the \tauhtauh channel. Additionally, the visible mass of the $\PGt\PGt$ system must be less than 125\GeV to ensure that the $\PGt\PGt$ system is compatible with an SM Higgs boson. To minimize diboson and \wjet contributions, the two \PGt candidates must pass a loose criterion of $\Delta R_{\PGt\PGt}<2.0$. Furthermore, a selection on the total transverse mass (\mttot) greater than 100\GeV is required. The \mttot distribution is used to extract the signal for this analysis and is defined as:
\begin{equation}
    \mttot = \sqrt{
        (E_\mathrm{T}^{\PGt_1}+E_\mathrm{T}^{\PGt_2}+\ptmiss)^2
        - (p_x^{\PGt_1}+p_x^{\PGt_2}+p_x^{\text{miss}})^2
        -(p_y^{\PGt_1}+p_y^{\PGt_2}+p_y^{\text{miss}})^2,
    }
    \label{eqn:mttot}
\end{equation}
where $p_{x}^{\text{miss}}$ and $p_{y}^{\text{miss}}$ are the magnitudes of the $x$ and $y$ components of \ptvecmiss. Here $\PGt_1$ and $\PGt_2$ are the leading and the subleading electrons, muons, or \tauh candidates in all three channels.
This selection reduces the background contributions from the Drell--Yan backgrounds in the leptonic channels and is chosen to maximize the expected significance of the analysis. 

During the 2017 data taking, a gradual shift in the timing of the inputs of the ECAL L1 trigger in the region at $\abs{\eta}> 2.0$ caused a specific trigger inefficiency. For events containing in this region an electron or a photon with $\pt \ge$ 50\GeV, or a jet with $\pt \ge$ 100\GeV, the efficiency loss is up to $\approx10-20\%$, depending on $\pt$, $\eta$, and time~\cite{Sirunyan:2020zal}. Correction factors are computed from data and applied to the acceptance evaluated from simulation for the 2017 samples. Finally, a section of the HCAL was not functioning during a part of the 2018 data-taking period corresponding to 65\% of the total integrated luminosity recorded in that year, leading to irrecoverable mismeasurements in a localized region of the detector ($-1.57<\phi< -0.87, -3.0 < \eta < -1.3$). To avoid contamination from such mismeasurements, events where any jet with $\pt > 30\GeV$ is found in the corresponding $\eta$-$\phi$ region are rejected in the analysis of the corresponding partial 2018 data set.

On top of JES corrections to \ptvecmiss, this analysis adopts an additional correction similar to that in Ref.~\cite{Sirunyan:2017khh}. The differences between data and simulation are applied as corrections to the direction and magnitude of the \ptvecmiss in simulated events. Simulated $\PZ \to \PGm\PGm$ events are used to estimate the hadronic recoil, which is then compared with selected $\PZ \to \PGm\PGm$ events in data. This correction to \ptvecmiss is applied to \zjet events. More details on this method can be found in Ref.~\cite{Sirunyan:2019kia}. Additionally, for simulated \ttbar events, a top quark \pt reweighing is applied to match the \pt distribution observed in data~\cite{CMS:2018htd}.

\subsection{Signal extraction and background estimation}\label{sec:bkgestimation}

Backgrounds are estimated using a combination of simulated samples and control samples in data. The major backgrounds arising from jets misidentified as $\tauh$ candidates, including events from \wjet, QCD multijet, and \ttbar production, are estimated using data, with a so-called ``misidentification factor method''. The other backgrounds, from \zjet, SM Higgs boson, single top quark, diboson production processes, and the background from \ttbar production with two real $\tau$ leptons, are estimated from simulation. The measurements of the cross sections for Drell--Yan~\cite{CMS:2018lkr} and single top quark~\cite{CMS:2016lel} processes at the LHC are consistent with the NNLO theoretical predictions, within the quoted uncertainties, supporting the use of theory-based normalizations for these backgrounds.
 
The background with jets misidentified as $\tauh$ candidates, called the ``mis-ID background'', is obtained with a method similar to that in Ref.~\cite{CMS:2022kdi}. The method is based on measuring two quantities: the number of events where jets pass the nominal $\tauh$ identification (ID) criteria (called isolated), corresponding to the signal region (SR) selection of the corresponding analysis, and the number of events where jets satisfy a set of loose ID criteria, but fail the nominal $\tauh$ ID criteria (referred to as ``antiisolated'' in the following).

The MF is defined as the ratio of the two numbers:
\begin{equation}
\mathrm{MF} = \frac{\mbox{number of events passing preselection and Medium $D_\text{jet}$ ID discriminant}}{\mbox{number of events passing preselection and failing the Medium $D_\text{jet}$ ID discriminant}}
\label{eq:jetToTauFakeRate}
\end{equation}
and is applied as a weight to events that pass the corresponding SR selection except that the events are required to select $\tauh$ candidates satisfying the loose preselection criteria, but failing the nominal $\tauh$ ID criteria (application region). The MF is estimated as a function of the jet multiplicity (0 or $\geq 1$), separately for the QCD multijet, $\PW$+jets, and $\ttbar$ backgrounds. 
A control region (CR) enriched in QCD multijet events is obtained by applying the same selection as used in the analysis, except that the two $\PGt$ leptons are required to be of the same sign (SS) instead of the opposite sign (OS). A CR enriched in \wjet events is obtained by applying the same selection as used in the analysis, except that the Higgs boson \pt, which is the visible \pt of the $\tau$ lepton pair, is required to be less than 65\GeV and \mT, which is the transverse mass of the lepton and the \ptmiss, is required to be greater than 70\GeV. Finally, a CR enriched in $\ttbar$ events is obtained by applying the same selection as used in the analysis, except that the number of b-tagged jets is required to be at least 1.

The following steps are performed successively to estimate the mis-ID background:
\begin{enumerate}
\item Measure the MFs as a function of the \pttauh: The MFs are derived as the ratio of isolated to antiisolated $\tauh$ candidates in QCD, \wjet, and $\ttbar$ events in dedicated CRs, and are fitted with a linear function.
\item Correct the MFs as a function of the \pt of the lepton that is not misidentified: The correction is derived by comparing the distributions of isolated events in the CR with antiisolated events in the CR, reweighted with the raw MFs derived in the previous step.
\item Correct the MFs for differences in selection between the CRs and the SR: The correction in the QCD CR (OS/SS correction), and the differences in the selection of the Higgs boson \pt and the \mT in the \wjet CR. The correction for the inversion of the Higgs boson \pt and \mT selections in the \wjet CR is determined through simulations by comparing the isolated events in the CR to the weighted antiisolated events as a function of \mT. The correction for the sign inversion in the QCD CR is computed using data. It is determined as a function of visible mass of the $\PGt\PGt$ system in a CR with antiisolated lepton i.e $0.1 < I_\text{rel}^{\Pe} < 0.2$ ($0.15 < I_\text{rel}^{\PGm} < 0.25$) in the \etau (\mutau) final state. The MF is computed by using step 1 and step 2 and is applied as a weight to OS events with an antiisolated tau and antiisolated lepton. These weighted events are then compared with OS events with anti-isolated lepton and an isolated tau to obtain the correction. 
\item Determine the relative fraction of QCD, \wjet, and \ttbar events: The \ttbar and \wjet yields are estimated using simulations, whereas the QCD yield is estimated by subtracting all the backgrounds from data in the antiisolated region.
\item Apply the MFs weighted by the corresponding fractions, to obtain the mis-ID background from the simulated events in the SR.
\end{enumerate}

\begin{figure}[!htb]
  \begin{center}
    \includegraphics[width=0.38\textwidth]{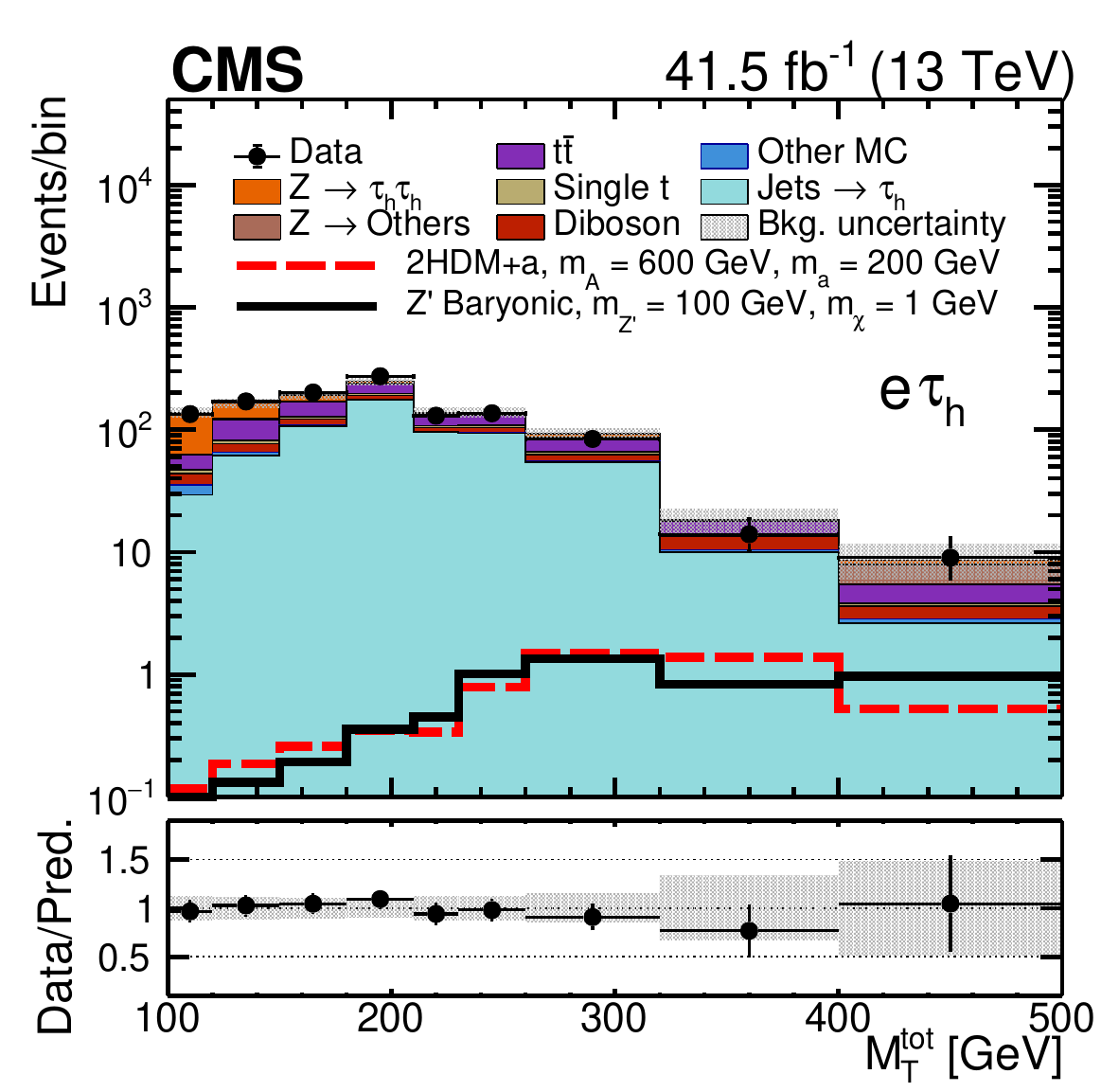}
    \includegraphics[width=0.38\textwidth]{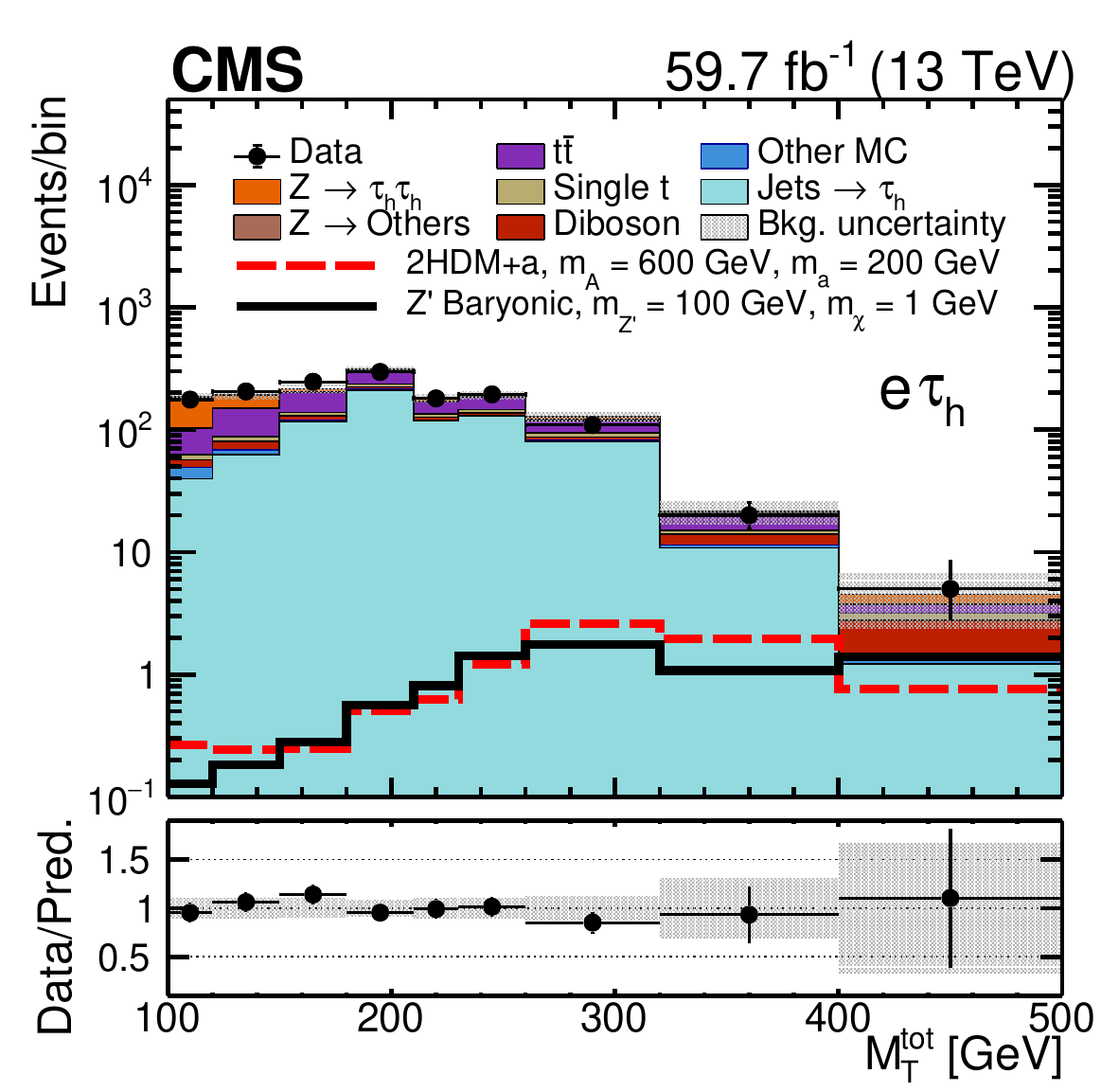}
     \includegraphics[width=0.38\textwidth]{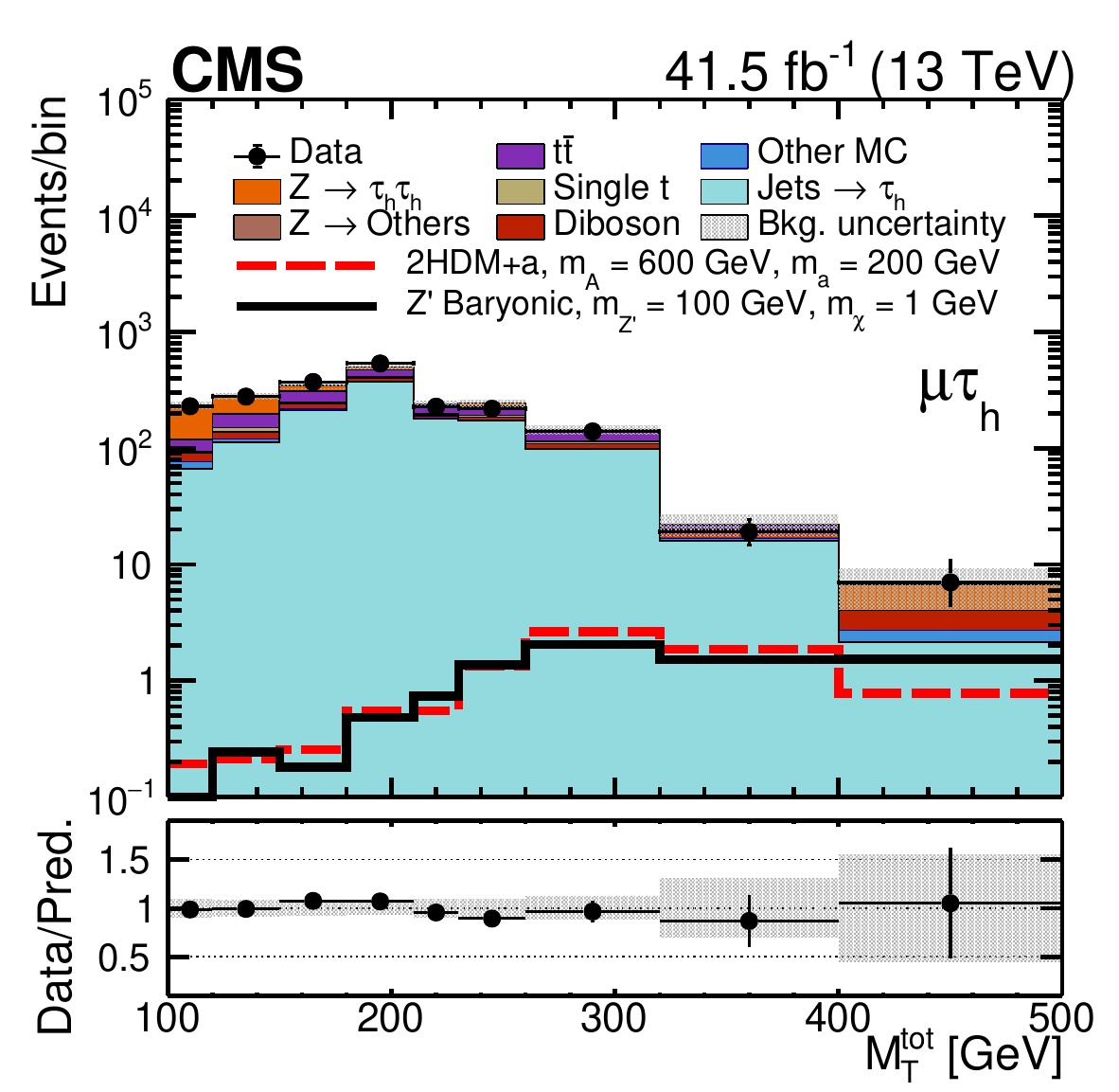}
    \includegraphics[width=0.38\textwidth]{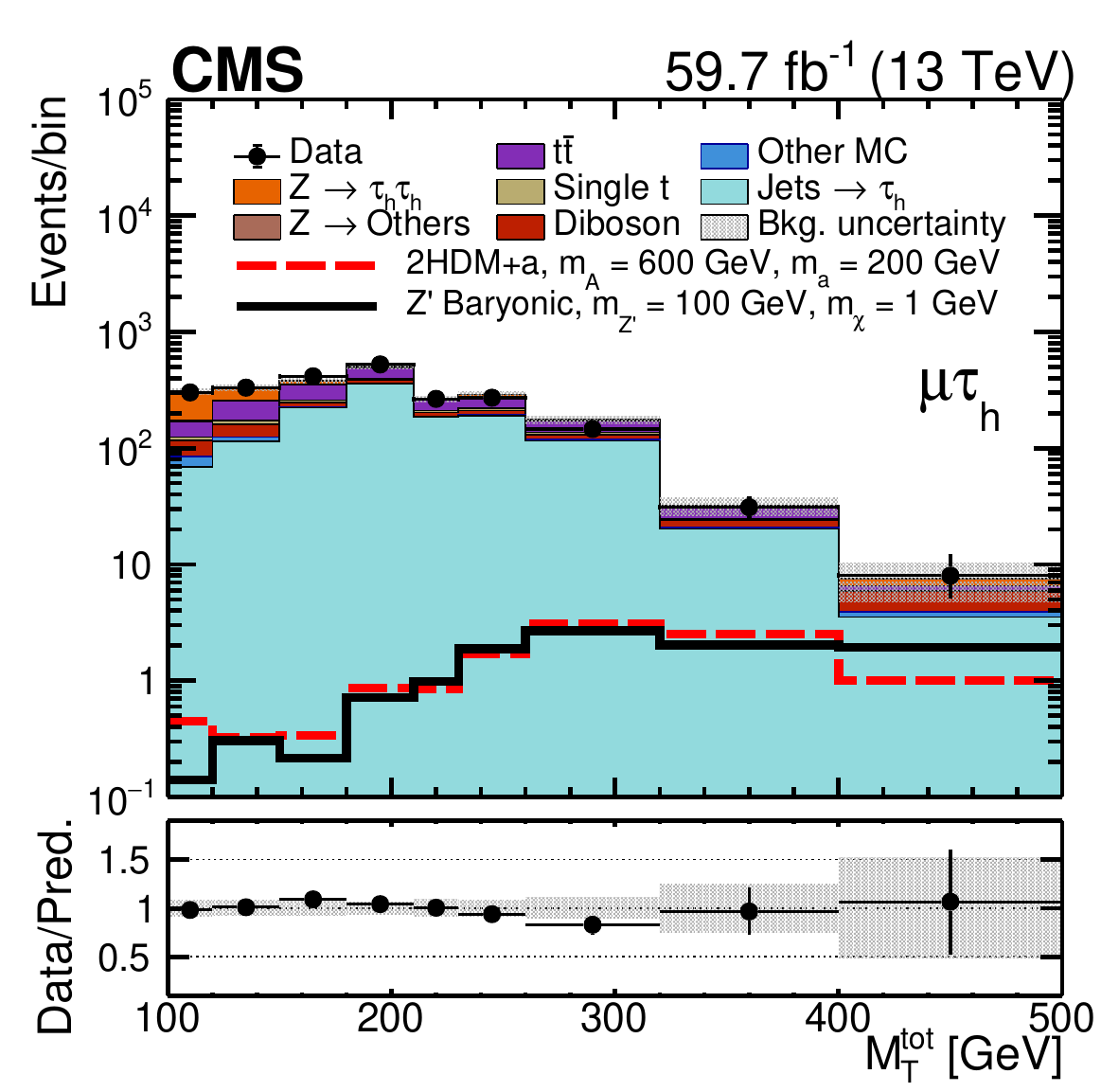}
     \includegraphics[width=0.38\textwidth]{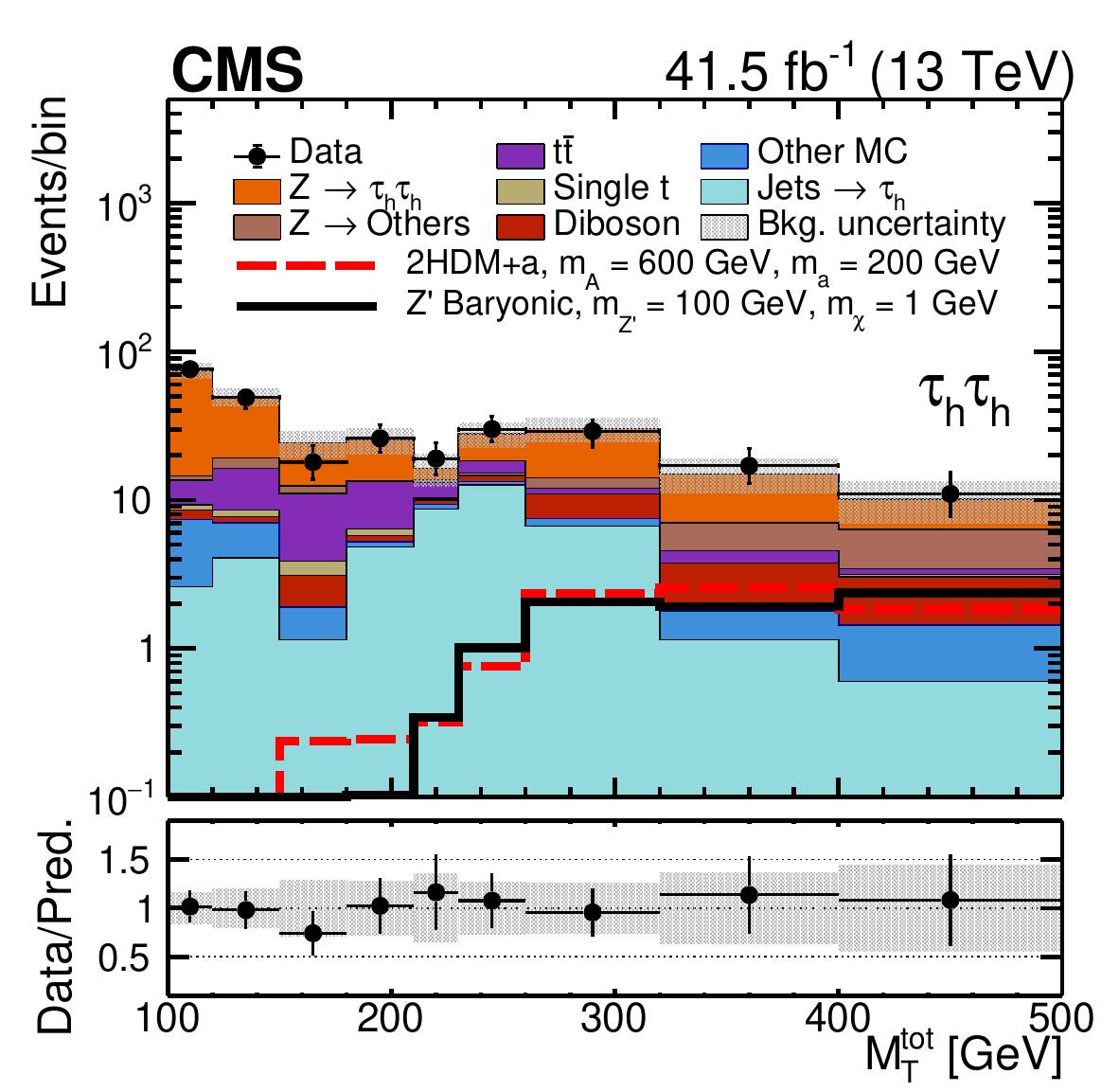}
    \includegraphics[width=0.38\textwidth]{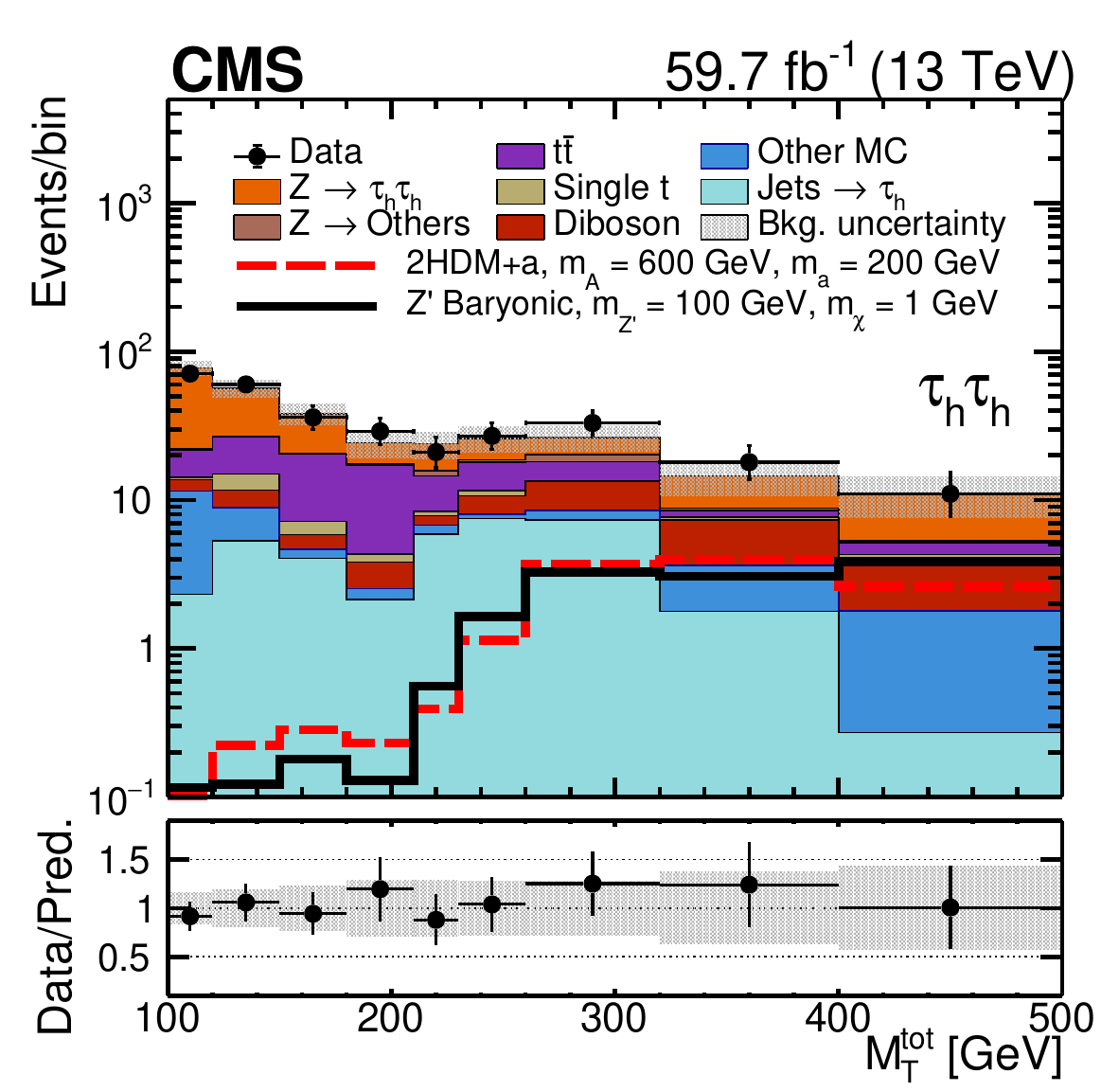}
  \end{center}
	\caption{Distributions of the total transverse mass \mttot in the SRs, for observed data and SM prediction in the \etau (upper), \mutau (center), and \tauhtauh (lower) final states in 2017 (left) and 2018 (right) after the simultaneous maximum likelihood fit. Representative signal distributions are shown for the \twohdma (dashed red curve) and \baryonicZprime (solid black curve) models. The data points are shown with their statistical uncertainties, and the last bin includes overflow. The ``Other MC'' background contribution includes events from ggh, VBF, $\PW\Ph$, $\PZ\Ph$, and electroweak vector boson production. The uncertainty band accounts for all systematic and statistical sources of uncertainty, after the fit to the data.
  }
  \label{fig:mt_postfit}
\end{figure}      
 
The signal is extracted with a binned maximum likelihood fit to the \mttot distributions (Eq. (\ref{eqn:mttot})) in the different channels for the SR.

Figure~\ref{fig:mt_postfit} shows the comparison of data and expected backgrounds for the \mttot distributions after the simultaneous likelihood fit to data in the SRs. Table~\ref{tab:event_yields_1718} shows the post-fit yields extracted from the \mttot distribution for the expected number of background events and the number of events observed in 2017 and 2018 data combined. The observed number of events is in good agreement with the estimated number of events by the SM backgrounds.

\begin{table}[htbp]
  \centering
  \caption{The post-fit estimated background yields and the observed number of events for the \mttot distribution in the SR, combined for the 2017 and 2018 data. The ``Other MC'' background contribution includes events from ggh, VBF, $\PW\Ph$, $\PZ\Ph$, \PZ $\rightarrow$ Others, and electroweak vector boson production. The uncertainties in the total expected yields include both statistical and systematic contributions.
    \label{tab:event_yields_1718}
  }
  \begin{tabular}{llll}
   Expected Backgrounds   &   \etau  & \mutau &  \tauhtauh \\
    \hline
    \PZ $\rightarrow$ \tauhtauh & $ 339 \pm 49 $  & $ 557 \pm 73 $  & $ 297 \pm 38 $ \\
    \ttbar & $ 542 \pm 48 $  & $ 744 \pm 61 $  & $ 99 \pm 10 $   \\
    Single Top & $ 94.9 \pm 8.6 $    & $ 122 \pm 10 $    & $ 11.8 \pm 1.2 $    \\
    Diboson  & $ 133 \pm 11 $    & $ 279 \pm 27 $  & $ 34.5 \pm 3.0 $   \\
    Other MC   & $ 64.4 \pm 6.5 $    & $ 84.4 \pm 7.0 $    & $ 50.7 \pm 4.8 $   \\
    Jets $\rightarrow$ \tauh & $ 1400 \pm 68 $  & $ 2508 \pm 95 $ & $ 79.1 \pm 5.9 $   \\ \\
    Total Background  & $ 2573 \pm 98 $ & $ 4294 \pm 138 $ & $ 572 \pm 40 $ \\ \\
    Data  & 2582 & 4311 & 581

  \end{tabular}
\end{table}

\section{Systematic uncertainties}\label{sec:sys_unc}
All systematic uncertainties affecting the analysis are summarized along in Table~\ref{tab:unc}, with their correlations between data sets from different years and different final states. These uncertainties are considered as nuisance parameters within the likelihood function. The selection efficiency correction factors are applied to the signal and all the background processes estimated through simulations, and uncertainties are included that account for differences between simulation and observation in the reconstruction and identification efficiencies for various particle candidates.

\begin{table}
\topcaption{Systematic uncertainties with their source, magnitude and correlations between data samples from either different data-taking years or different final states}
\label{tab:unc}
\centering
\resizebox{\textwidth}{!}
{%
\begin{tabular}{lccc}
Uncertainty & Magnitude & Correlation years & Correlation final states \\
\hline
	$\tauh$ identification & $\pt$-dependent & 0\% & 100 (0)\% for $\Pell\tauh$ ($\tauh\tauh$) \\
$\tauh$ ID, for high \pttauh & Event-dependent & 100\% & 100\% \\
$\tauh$ energy scale & 0.7--1.2\% & 0\% & 100\% \\
Electron misidentified as $\tauh$ & Event-dependent& 0\% & 0\% \\
Muon misidentified as $\tauh$ & Event-dependent & 0\% & 0\% \\
Electron/muon identification & 2\% & 0\% & 0\% \\
Energy scale for lepton mis-ID as  $\tauh$ & 1\% & 0\% & 0\% \\
Trigger & Event-dependent & 0\% & 0\% \\
Electron energy scale & Event-dependent & 100\% & 100\% \\
Muon energy scale & 0.4--2.7\% & 100\% & 100\% \\
b jet veto & Event-dependent & partial & 100\% \\
Integrated luminosity & 2--3\% & partial & 100\% \\
\ttbar cross section & 6\% & 100\% & 100\% \\
Diboson cross section & 5\% & 100\% & 100\% \\
Single top quark cross section & 5\% & 100\% & 100\% \\
Drell--Yan cross section & 4\% & 100\% & 100\% \\
L1 mistiming & Event-dependent & 0\% & 100\% \\
\PZ boson $\pt$ reweighting & 10\% & 100\% & 100\% \\
Top quark $\pt$ reweighting & Event-dependent & 100\% & 100\% \\
JES & Event-dependent & 100\% & 100\% \\
JER & Event-dependent & 100\% & 100\% \\
\ptvecmiss unclustered ES & Event-dependent & 0\% & 100\%\\
\ptvecmiss recoil corrections & Event-dependent & 0\% & 100\% \\
MF in $\Pell\tauh$ & Event-dependent & partial & 0\% \\
MF in \tauhtauh & Event-dependent & partial & 0\% \\
MF in \tauhtauh, for $\pttauh > 100\GeV$ & Event-dependent & 0\% & 0\% \\
\end{tabular}
}
\end{table}

The uncertainty in the \tauh identification efficiency is considered as a \pttauh- and decay-mode-dependent uncertainty ranging from 3 to 5\%~\cite{CMS:2022prd}. An additional shape-based uncertainty on the \mttot distribution in the range 10--15\% is added for \tauh candidates with $\pt > 100\GeV$.  The uncertainty in the \tauh trigger efficiency is also \pttauh-dependent and ranges from 4 to 9\%. Since both the \tauh identification and \tauh trigger uncertainties are statistically dominated, they are considered as uncorrelated across data-taking years. 

An uncertainty covers the misidentification of electrons or muons as \tauh candidates.  It is dependent on \pttauh, $\eta$, and the decay mode. It is also considered as uncorrelated between the samples of different data-taking years. The uncertainty in the \tauh energy scale is in the range 0.7--1.2\%, depending on \pttauh and on the \tauh decay mode. It is treated as uncorrelated between data-taking years and decay modes. An additional uncertainty of 1\% is also considered for the energy scale of the misidentified leptons and it is considered uncorrelated across the data-taking years.

The uncertainty in the electron and muon reconstruction, efficiency, and identification is 2\% for all years, uncorrelated across data-taking years~\cite{CMS:2020uim,CMS:2012nsv,CMS:2018rym}. 
The electron energy scale uncertainties are also derived event-by-event and are treated as shape uncertainties. Since they are systematically dominated~\cite{CMS:2020uim}, these uncertainties are treated as correlated across data-taking years. The uncertainty related to the muon energy scale is between 0.4 and 2.7\%, and correlated between data-taking years. A shape-based uncertainty depending on \pt and $\eta$ for the electron and muon triggers is considered. 	

For the background processes, cross section uncertainties of 5\% for diboson and single top quark~\cite{Kidonakis:2013zqa, Campbell:2011bn, Gehrmann:2014fva, CMS:2016lel}, 4\% for Drell--Yan~\cite{Melnikov:2006kv, CMS:2018lkr}, and 6\% for \ttbar backgrounds are also considered~\cite{Kidonakis:2013zqa,Czakon:2011xx}. An additional uncertainty of 1-2\% to account for the L1 trigger ECAL timing shift is also considered. The uncertainties related to the integrated luminosities for the data-taking period 2017-2018 amounts to 2.3-2.5\%, while overall uncertainty for total Run-2 period is 1.6\%~\cite{CMS:2018elu,CMS:2019jhq}. In simulated $\PZ$+jets events, a shape uncertainty of 10\% of the \PZ boson \pt reweighting correction, to account for higher-order effects, is used. A shape-based uncertainty is applied to the \ttbar process to correct the top quark \pt distribution. For the Drell--Yan process, where \ptvecmiss recoil corrections are applied, an uncertainty in the calibration and resolution of the hadronic recoil are applied~\cite{Sirunyan:2019kia}.  For all other background processes estimated from simulation, an uncertainty in \ptvecmiss related to unclustered energy is taken into account. The \ptvecmiss uncertainties are considered as uncorrelated across data-taking years. Uncertainties in the JES are included on an event-by-event basis and considered as correlated across data-taking years. 

For the background with jets misidentified as \tauh candidates, different sources of uncertainty are considered. The MFs are parameterized linearly as function of \pttauh. Two uncertainties per fit function are considered. They are obtained by uncorrelating the fit uncertainties in the slope and intercept of the fit functions obtained in the three CRs, i.e, the \wjet, QCD multijet, and \ttbar CRs. A shape-based uncertainty for the lepton $\pt$ correction for all three CRs is considered. An additional shape-based uncertainty to cover the OS/SS correction in the QCD CR is applied. A shape-based uncertainty is considered for \tauh candidates with $\pt > 100 \GeV$, following Ref.~\cite{CMS:2024jbe}.

\section{Results}\label{sec:results}

The results are extracted from a maximum likelihood fit, using the CMS statistical analysis tool \textsc{Combine}~\cite{CMS:2024onh}. No significant excess of data above the SM background expectation is observed. Interpretations in the \twohdma and \baryonicZprime models are provided.

\begin{figure}[!htb]
  \begin{center}
    \includegraphics[width=0.45\textwidth]{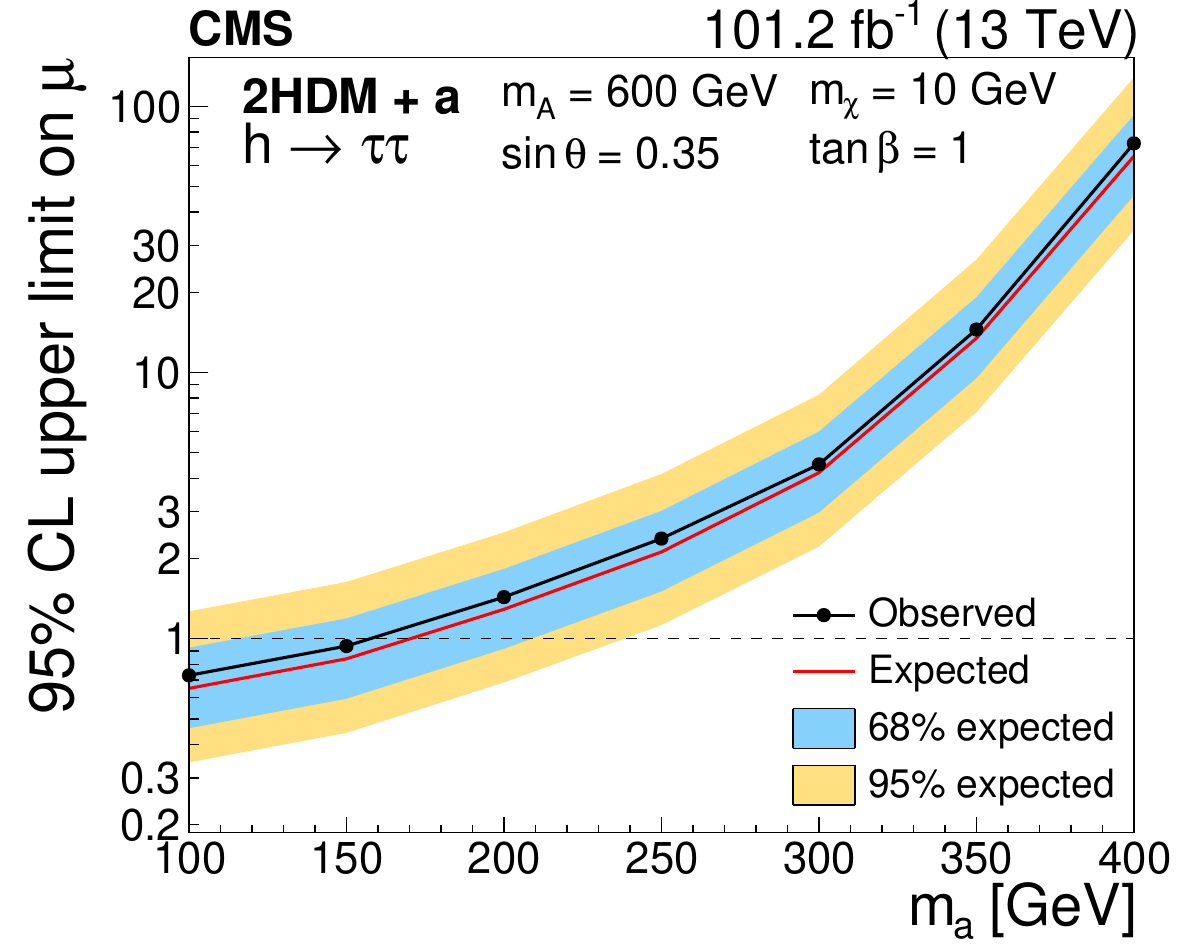}
    \includegraphics[width=0.45\textwidth]{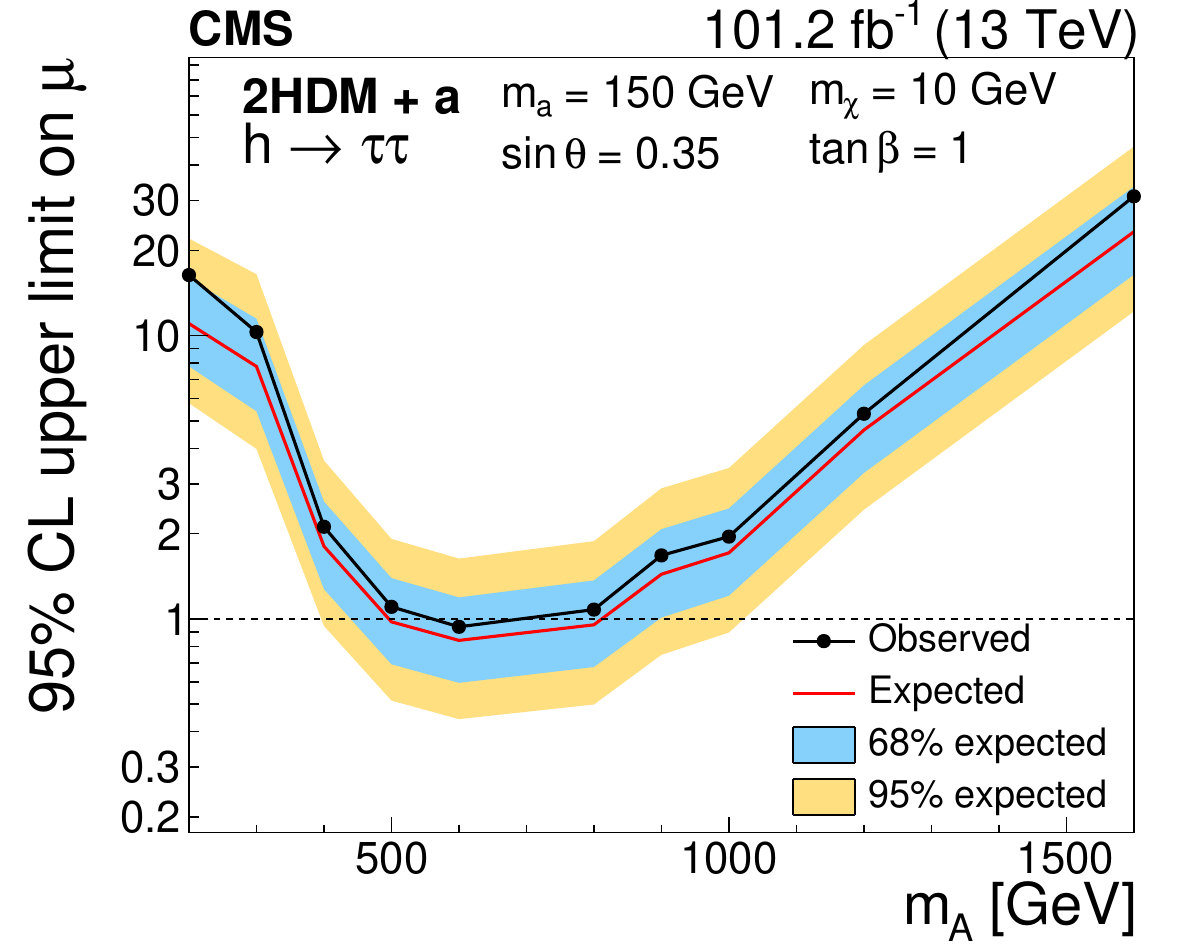}
    \includegraphics[width=0.45\textwidth]{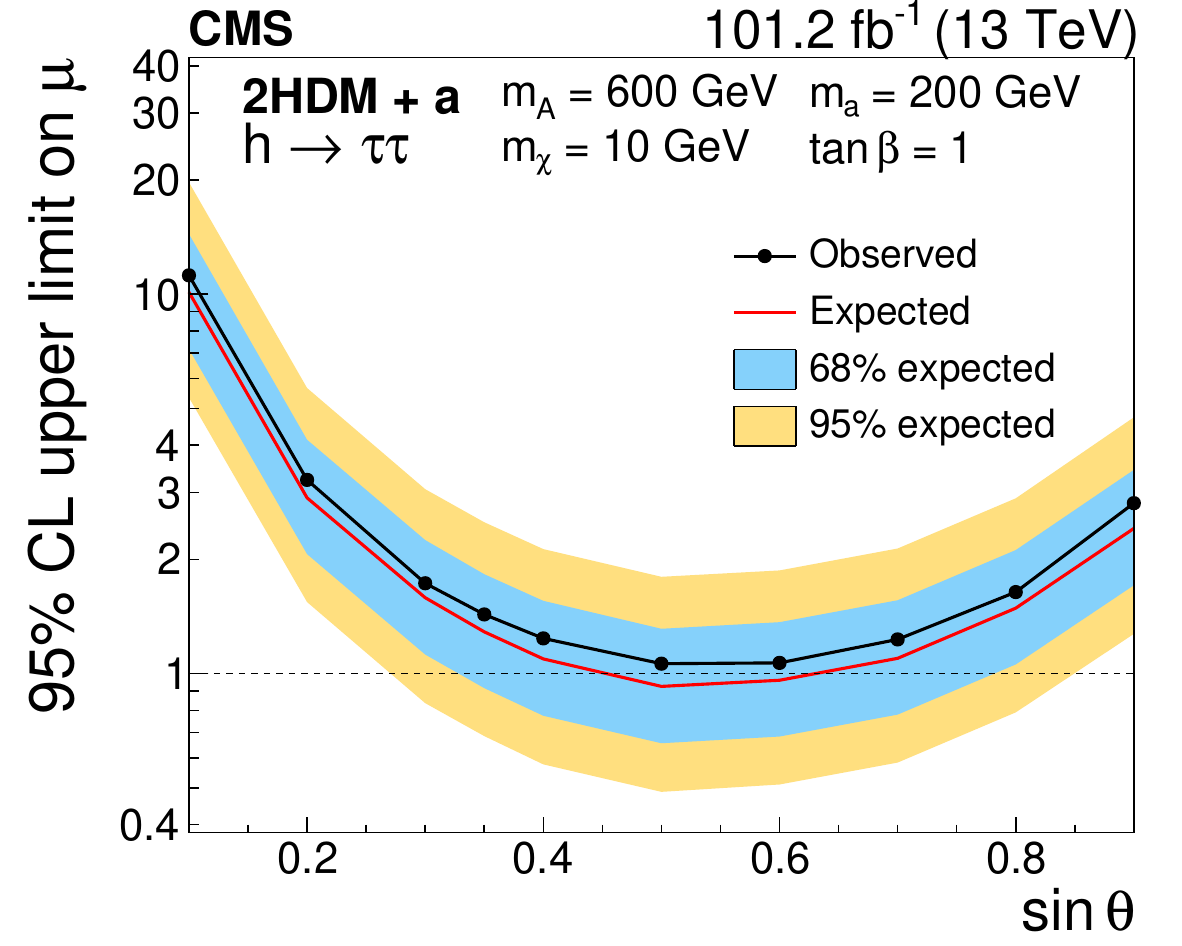}
    \includegraphics[width=0.45\textwidth]{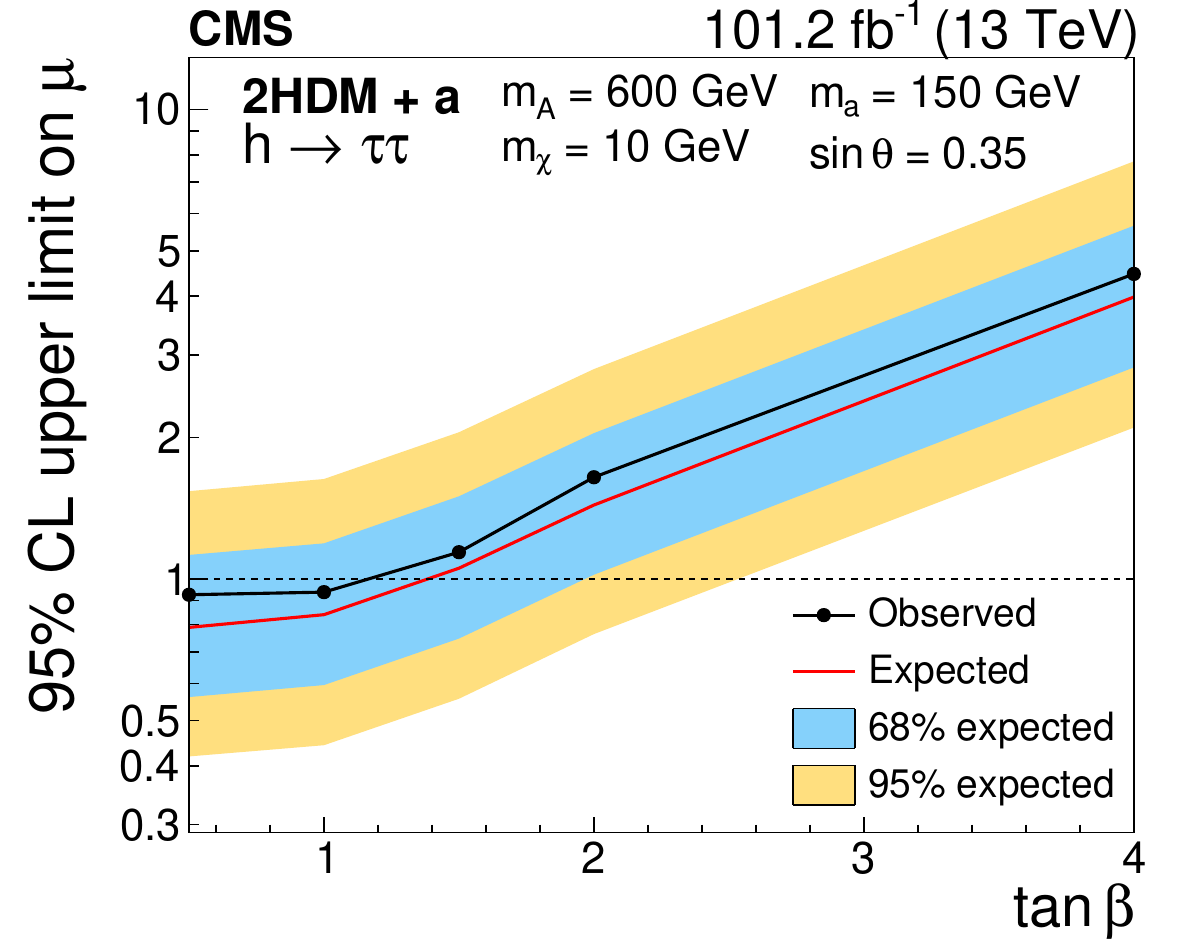}
  \end{center}
  \caption{The 95\% CL upper limits on $\mu$ for the \twohdma model as a function of \ma (upper let), \mA (upper right), \sintheta (lower left) and \tanb (lower right). The values assumed for the other parameters are shown on each figure. The interpolation between the points is linear.}
  \label{fig:twohdma_1d}
\end{figure}

The signal upper limits are determined with the signal strength modifier $\mu = \sigma/\sigma_{\text{theory}}$, where $\sigma_{\text{theory}}$ is the predicted production cross section of the particular model considered, calculated using the parameter choices described in Section~\ref{sec:intro}, and $\sigma$ corresponds to the upper limit on the observed cross section.

Upper limits are obtained using the data set collected in 2017--2018. They are calculated at 95\% confidence level (CL) using the CLs criterion~\cite{Junk:1999kv, Read:2002hq} computed with an asymptotic approximation~\cite{Cowan:2010js} by using the profile likelihood ratio modified for upper limits. 

For the \twohdma model, the one-dimensional upper limits on $\mu$ as a function of the lighter pseudoscalar mass \ma are shown in Fig.~\ref{fig:twohdma_1d} (upper left), using \sintheta = 0.35, \mA = 600\GeV, \tanb = 1, and \mchi = 10\GeV.  Similar limits as functions of \mA are shown in Fig.~\ref{fig:twohdma_1d} (upper right), using \sintheta = 0.35, \ma = 150\GeV, \tanb = 1, and \mchi = 10\GeV. 
Upper limits on $\mu$ as a function of \sintheta (\tanb) are shown in Fig.~\ref{fig:twohdma_1d} lower left (lower right), using \ma = 200\GeV, \mA = 600\GeV, \tanb = 1, and \mchi = 10\GeV (\ma = 150\GeV,  \mA = 600\GeV, \sintheta = 0.35, and \mchi = 10\GeV).

\begin{figure}[!htb]
  \begin{center}
     \includegraphics[width=0.55\textwidth]{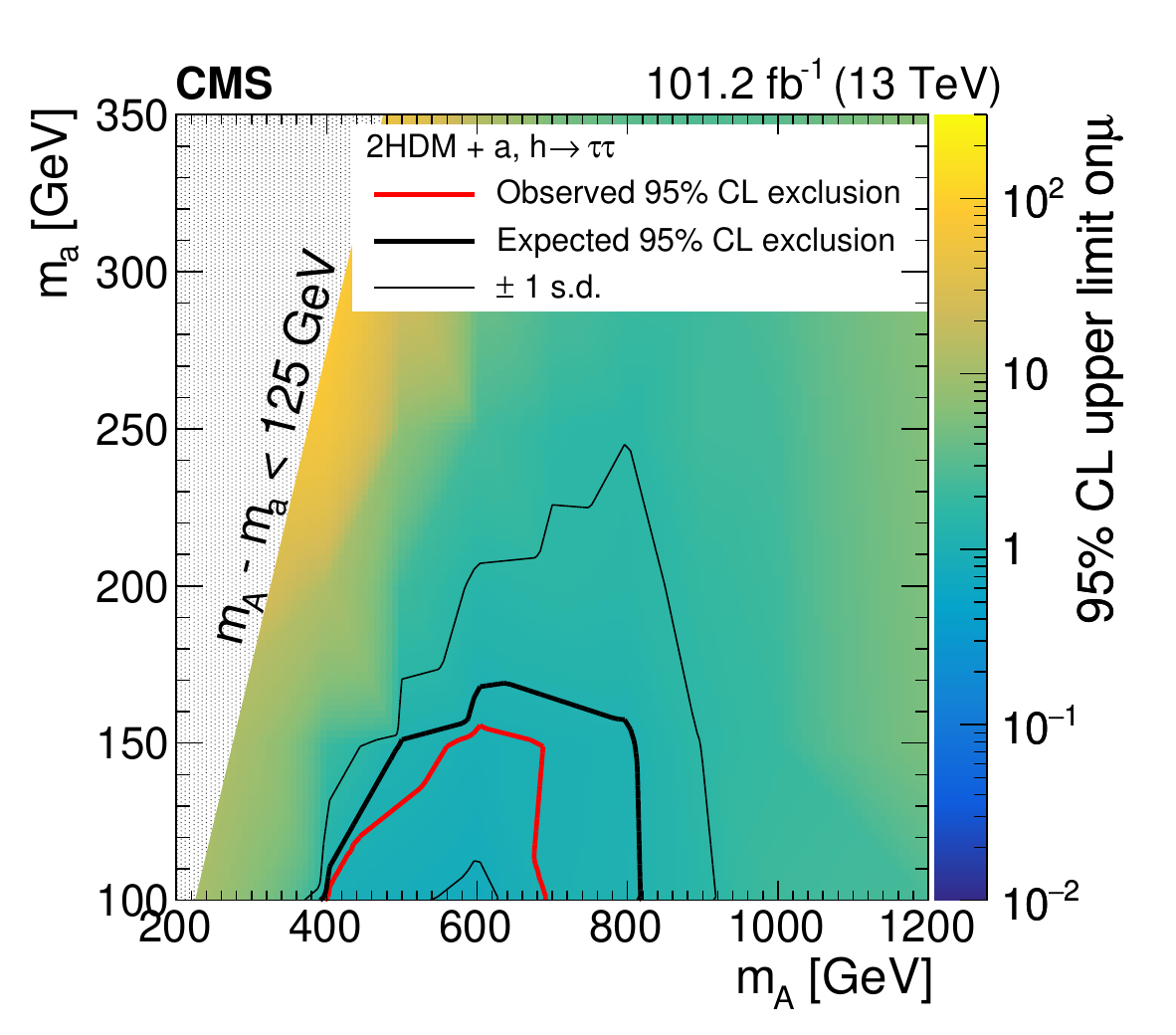}
     \end{center}
  \caption{The 95\% CL upper limits on $\mu$ in the (\mA,\ma) plane for the \twohdma model. The regions inside the red and black curves correspond to the observed and expected exclusions at 95\% CL, respectively.}
  \label{fig:twohdma_2d}
\end{figure}

Two-dimensional limits in the (\mA,\ma) plane, using \sintheta = 0.35, \tanb = 1, and \mchi = 10\GeV are shown in Fig.~\ref{fig:twohdma_2d}. Masses of \mA between 400 and 700 \GeV are excluded for small values of \ma ($\ma\approx 100\GeV$). 

\begin{figure}[!htb]
  \begin{center}
    \includegraphics[width=0.47\textwidth,height=6.5cm]{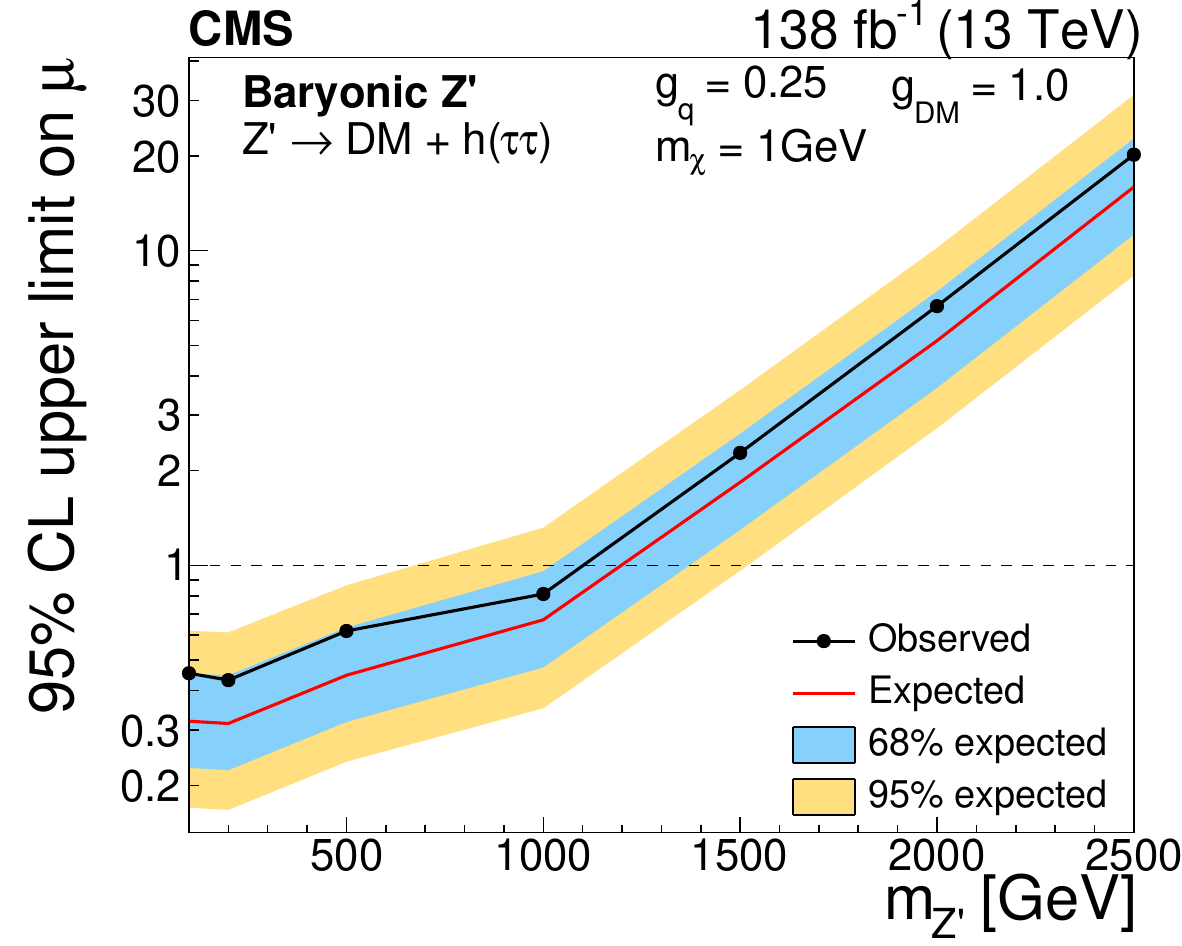}
    \includegraphics[width=0.49\textwidth]{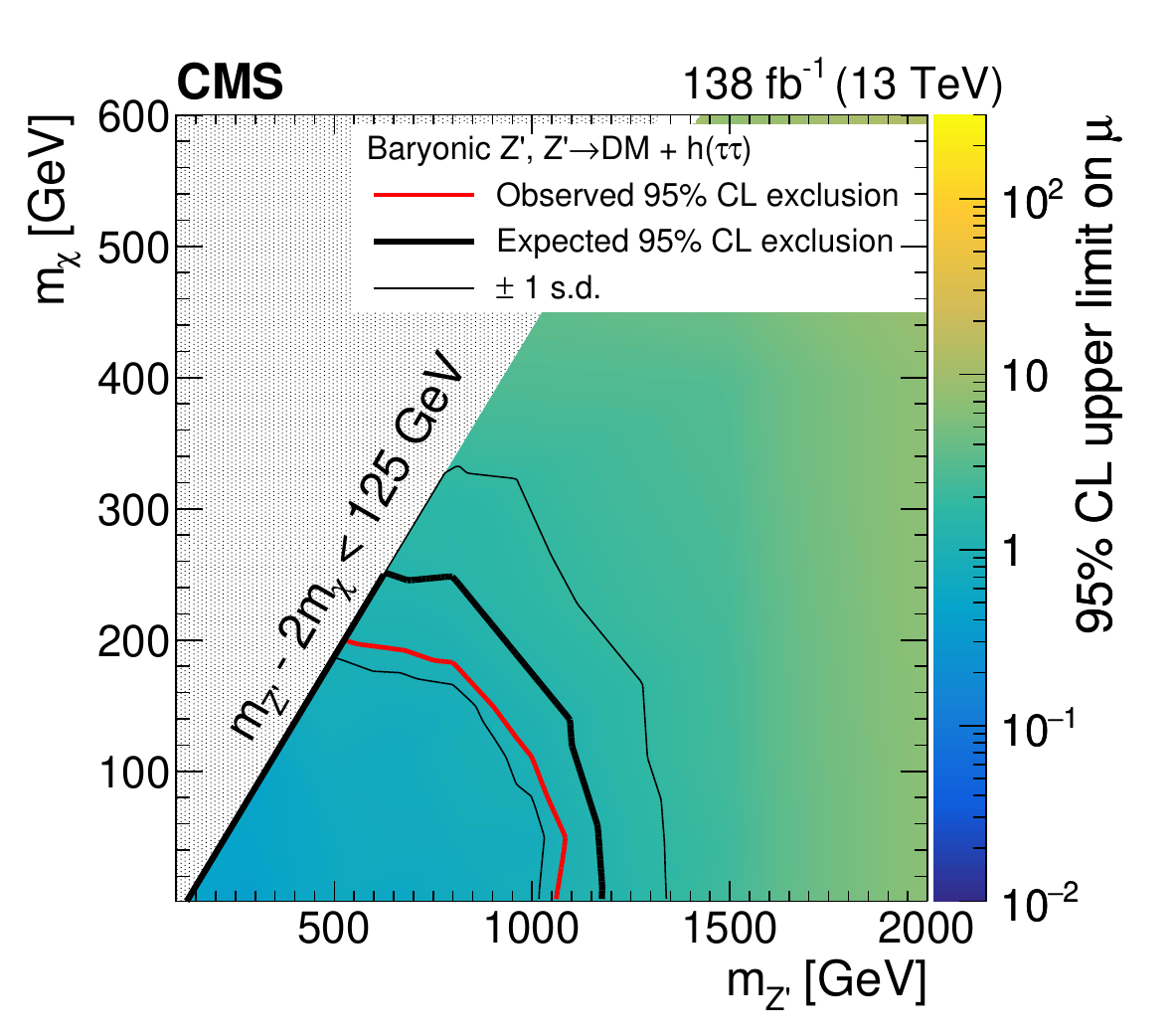}
  \end{center}
  \caption{The 95\% CL upper limits on $\mu$ for the \baryonicZprime model. Left: Observed and expected 95\% CL upper limits on $\mu$ as a function of \mZpr, using \mchi = 1\GeV. Right: 95\% CL upper limits on $\mu$ in the (\mZpr,\mchi) plane. The regions inside the red and black curve correspond to the observed and expected exclusions at 95\% CL, respectively. 
  }
  \label{fig:zprime}
\end{figure}

The one-dimensional upper limits on $\mu$ for the \baryonicZprime model, as a function of \mZpr are shown in Fig.~\ref{fig:zprime} (left), using 138\fbinv of data collected in 2016--2018. The two-dimensional limits on $\mu$ in the (\mZpr,\mchi) plane are shown in Fig.~\ref{fig:zprime} (right). In the previous result based on 2016 data, the observed exclusion region for the mediator mass reached 450\GeV for a DM mass of 1\GeV. After combination with the search based on 2017 and 2018 data presented in this paper, the observed (expected) exclusion region for \mZpr extends to around 1050 (1150)\GeV for the same DM mass.

\section{Summary}\label{sec:summary}

A search for dark matter produced in association with a Higgs boson decaying to a pair of \PGt leptons has been performed using a data set of proton-proton collisions at a center-of-mass energy of 13\TeV, corresponding to an integrated luminosity of 101\fbinv. The results are interpreted within the framework of two benchmark simplified models: the \twohdma model, where a heavy pseudoscalar couples to a Higgs boson and a lighter pseudoscalar that decays to dark matter particles, and the \baryonicZprime model, where a high mass resonance (\zp) decays into a pair of dark matter particles and a standard model Higgs boson. Upper limits at the 95\% confidence level are set on the product of the production cross section and branching fraction for both models.
In the \twohdma model, heavy pseudoscalar masses between 400 and 700\GeV are excluded for a light pseudoscalar mass around 100\GeV. For the \baryonicZprime model, the results are combined with those of an earlier search using an independent data set collected at the same center-of-mass energy, corresponding to an integrated luminosity of 36\fbinv. \PZpr masses up to 1050\GeV are excluded for a dark matter particle mass of 1\GeV, based on a data sample corresponding to a total integrated luminosity of 138\fbinv.

\begin{acknowledgments}
We congratulate our colleagues in the CERN accelerator departments for the excellent performance of the LHC and thank the technical and administrative staffs at CERN and at other CMS institutes for their contributions to the success of the CMS effort. In addition, we gratefully acknowledge the computing centers and personnel of the Worldwide LHC Computing Grid and other centers for delivering so effectively the computing infrastructure essential to our analyses. Finally, we acknowledge the enduring support for the construction and operation of the LHC, the CMS detector, and the supporting computing infrastructure provided by the following funding agencies: SC (Armenia), BMBWF and FWF (Austria); FNRS and FWO (Belgium); CNPq, CAPES, FAPERJ, FAPERGS, and FAPESP (Brazil); MES and BNSF (Bulgaria); CERN; CAS, MoST, and NSFC (China); MINCIENCIAS (Colombia); MSES and CSF (Croatia); RIF (Cyprus); SENESCYT (Ecuador); ERC PRG, RVTT3 and MoER TK202 (Estonia); Academy of Finland, MEC, and HIP (Finland); CEA and CNRS/IN2P3 (France); SRNSF (Georgia); BMBF, DFG, and HGF (Germany); GSRI (Greece); NKFIH (Hungary); DAE and DST (India); IPM (Iran); SFI (Ireland); INFN (Italy); MSIP and NRF (Republic of Korea); MES (Latvia); LMTLT (Lithuania); MOE and UM (Malaysia); BUAP, CINVESTAV, CONACYT, LNS, SEP, and UASLP-FAI (Mexico); MOS (Montenegro); MBIE (New Zealand); PAEC (Pakistan); MES and NSC (Poland); FCT (Portugal); MESTD (Serbia); MICIU/AEI and PCTI (Spain); MOSTR (Sri Lanka); Swiss Funding Agencies (Switzerland); MST (Taipei); MHESI and NSTDA (Thailand); TUBITAK and TENMAK (Turkey); NASU (Ukraine); STFC (United Kingdom); DOE and NSF (USA).

\hyphenation{Rachada-pisek} Individuals have received support from the Marie-Curie program and the European Research Council and Horizon 2020 Grant, contract Nos.\ 675440, 724704, 752730, 758316, 765710, 824093, 101115353, 101002207, 101001205, and COST Action CA16108 (European Union); the Leventis Foundation; the Alfred P.\ Sloan Foundation; the Alexander von Humboldt Foundation; the Science Committee, project no. 22rl-037 (Armenia); the Fonds pour la Formation \`a la Recherche dans l'Industrie et dans l'Agriculture (FRIA-Belgium); the Beijing Municipal Science \& Technology Commission, No. Z191100007219010, the Fundamental Research Funds for the Central Universities, the Ministry of Science and Technology of China under Grant No. 2023YFA1605804, and the Natural Science Foundation of China under Grant No. 12061141002 (China); the Ministry of Education, Youth and Sports (MEYS) of the Czech Republic; the Shota Rustaveli National Science Foundation, grant FR-22-985 (Georgia); the Deutsche Forschungsgemeinschaft (DFG), among others, under Germany's Excellence Strategy -- EXC 2121 ``Quantum Universe" -- 390833306, and under project number 400140256 - GRK2497; the Hellenic Foundation for Research and Innovation (HFRI), Project Number 2288 (Greece); the Hungarian Academy of Sciences, the New National Excellence Program - \'UNKP, the NKFIH research grants K 131991, K 133046, K 138136, K 143460, K 143477, K 146913, K 146914, K 147048, 2020-2.2.1-ED-2021-00181, TKP2021-NKTA-64, and 2021-4.1.2-NEMZ\_KI-2024-00036 (Hungary); the Council of Science and Industrial Research, India; ICSC -- National Research Center for High Performance Computing, Big Data and Quantum Computing and FAIR -- Future Artificial Intelligence Research, funded by the NextGenerationEU program (Italy); the Latvian Council of Science; the Ministry of Education and Science, project no. 2022/WK/14, and the National Science Center, contracts Opus 2021/41/B/ST2/01369 and 2021/43/B/ST2/01552 (Poland); the Funda\c{c}\~ao para a Ci\^encia e a Tecnologia, grant CEECIND/01334/2018 (Portugal); the National Priorities Research Program by Qatar National Research Fund;  MICIU/AEI/10.13039/501100011033, ERDF/EU, "European Union NextGenerationEU/PRTR", and Programa Severo Ochoa del Principado de Asturias (Spain); the Chulalongkorn Academic into Its 2nd Century Project Advancement Project, and the National Science, Research and Innovation Fund via the Program Management Unit for Human Resources \& Institutional Development, Research and Innovation, grant B39G670016 (Thailand); the Kavli Foundation; the Nvidia Corporation; the SuperMicro Corporation; the Welch Foundation, contract C-1845; and the Weston Havens Foundation (USA).
\end{acknowledgments}

\bibliography{auto_generated}

\cleardoublepage \appendix\section{The CMS Collaboration \label{app:collab}}\begin{sloppypar}\hyphenpenalty=5000\widowpenalty=500\clubpenalty=5000\input{SUS-23-012-public-authorlist.tex}\end{sloppypar}
\end{document}

%% file: SUS-23-012-public-authorlist.tex
\cmsinstitute{Yerevan Physics Institute, Yerevan, Armenia}
{\tolerance=6000
V.~Chekhovsky, A.~Hayrapetyan, V.~Makarenko\cmsorcid{0000-0002-8406-8605}, A.~Tumasyan\cmsAuthorMark{1}\cmsorcid{0009-0000-0684-6742}
\par}
\cmsinstitute{Institut f\"{u}r Hochenergiephysik, Vienna, Austria}
{\tolerance=6000
W.~Adam\cmsorcid{0000-0001-9099-4341}, J.W.~Andrejkovic, L.~Benato\cmsorcid{0000-0001-5135-7489}, T.~Bergauer\cmsorcid{0000-0002-5786-0293}, S.~Chatterjee\cmsorcid{0000-0003-2660-0349}, K.~Damanakis\cmsorcid{0000-0001-5389-2872}, M.~Dragicevic\cmsorcid{0000-0003-1967-6783}, P.S.~Hussain\cmsorcid{0000-0002-4825-5278}, M.~Jeitler\cmsAuthorMark{2}\cmsorcid{0000-0002-5141-9560}, N.~Krammer\cmsorcid{0000-0002-0548-0985}, A.~Li\cmsorcid{0000-0002-4547-116X}, D.~Liko\cmsorcid{0000-0002-3380-473X}, I.~Mikulec\cmsorcid{0000-0003-0385-2746}, J.~Schieck\cmsAuthorMark{2}\cmsorcid{0000-0002-1058-8093}, R.~Sch\"{o}fbeck\cmsAuthorMark{2}\cmsorcid{0000-0002-2332-8784}, D.~Schwarz\cmsorcid{0000-0002-3821-7331}, M.~Sonawane\cmsorcid{0000-0003-0510-7010}, W.~Waltenberger\cmsorcid{0000-0002-6215-7228}, C.-E.~Wulz\cmsAuthorMark{2}\cmsorcid{0000-0001-9226-5812}
% ...
\par}
\vskip\cmsinstskip
\dag:~Deceased\\
$^{1}$Also at Yerevan State University, Yerevan, Armenia\\
$^{2}$Also at TU Wien, Vienna, Austria\\
% ...
